\documentclass[11pt,fleqn]{elsarticle}
\usepackage{t1enc}
\usepackage{theorem}
\usepackage{amsmath,amssymb}
\usepackage{upref}
\usepackage{lineno}
\usepackage{pstricks}
\psset{unit=0.6}
\newpsobject{showgrid}{psgrid}{%
  subgriddiv=1,griddots=10,gridlabels=6pt}
\usepackage{calc}
\newcounter{hours}\newcounter{minutes}
\newcommand\printtime{\setcounter{hours}{\time/60}%
  \setcounter{minutes}{\time-\value{hours}*60}%
  \thehours\,h\,\theminutes}
\newcommand\dateandtime{\today\quad\printtime}
\def\QED{\relax\ifmmode\eqno{\hbox{\petitcarre}}\else
{\unskip\nobreak\hfil\penalty50
   \hskip2em\hbox{}\nobreak\hfil
  $\Box$
   \parfillskip=0pt \finalhyphendemerits=0
    \par\smallskip}\fi}

\def\bb{{\blacktriangleright}}
\def\ff{{\blacktriangleleft}}

\newcommand\e{\varepsilon}
\newtheorem{theo}{Theorem} 
\newtheorem{prop}[theo]{Proposition}
\newtheorem{lem}[theo]{Lemma}

{\theorembodyfont{\normalfont}%
  \newtheorem{example}[theo]{Example}
  \newtheorem{rema}[theo]{Remark}}

\newenvironment{preuve}{\noindent {\bf Proof }}{\QED}

\newenvironment{sketch}{\noindent {\bf Sketch of proof }}{\QED}

\newcommand{\Ins}{\mathop{\mathrm{Ins}}}
\newcommand\ins[2]{{}_{#1}B_{#2}}
\renewcommand\ins[2]{{}_{#1\!}T_{\!#2}}
\renewcommand\ins[2]{{}^{#1\!}B^{#2}}
\newcommand\wrd[2]{{}_{#1}W_{#2}}
\renewcommand\wrd[2]{{}_{#1\!}W_{\! #2}}
\newcommand{\sing}[1]{V_#1}
\newcommand{\wrdb}[2]{{}_{#1}W_{#2} }
\renewcommand{\wrdb}[2]{{}_{#1\!}W_{\! #2} }
\newcommand{\singb}[1]{V_#1}
\renewcommand{\singb}[1]{V_{\! #1}}
\newcommand{\wrdi}[2]{{}_{#1}I_{#2} }
\newcommand{\singi}[1]{I_#1}

\newcommand{\rulc}[4]{\lbrack #1 {-} #2{\mid}#3  {-} #4  \rbrack} 
\newcommand{\langconc}[1]{\mathop{\cal K}(#1)} 
\newcommand{\produconc}[4]{#2,#3\, {\models}_{#1} #4} 

\newcommand{\produins}[4]{#2,#3\, {\Vdash}_{#1} #4} 
\newcommand{\ruli}[4]{\langle\langle  #1 {\mid} #3{-}#4  {\mid} #2
  \rangle\rangle} 

\newcommand\rul[4]{\langle  #1 {\mid} #3{-}#4  {\mid} #2  \rangle} 
\newcommand\Rul[4]{\langle  #1 {\mid} #2{-}#3  {\mid} #4  \rangle} 
\newcommand\Crul[4]{#1\##2\$#3\##4}
\newcommand{\produc}[4]{#2,#3\, \vdash_{#1} #4} 
\newcommand{\langf}[1]{{\cal F}({#1})} 
\newcommand{\langc}[1]{{\cal C}(#1)} 

\newcommand{\cir}[1]{{}^\sim #1}
\newcommand{\lin}[1] {Lin(#1)} 

\newcommand{\sublang}[2]{{L}_{#1}(#2)} 
\newcommand{\sublangprime}[2]{{L'}_{#1}(#2)}
\newcommand{\lang}[1]{{L}_{#1}} 
\newcommand{\langprime}[1]{{L'}_{#1}}
\usepackage[hypertex,hyperindex,pagebackref,final]{hyperref}
\numberwithin{theo}{section}
\numberwithin{equation}{section}

\begin{document}

\begin{frontmatter}
  

\title{Splicing systems and the Chomsky hierarchy\tnoteref{thanks}}

\author[igm]{Jean Berstel}
\author[liafa]{Luc Boasson}
\author[igm,p7]{Isabelle Fagnot\corref{cor}}

\tnotetext[thanks]{ Part of this work has been done during a
  sabbatical leave of the third author from University Paris 7, and
  during a stay at DIA (Dipartimento di Informatica e Applicazioni) at
  Università di Salerno, Italy}
\cortext[cor]{Corresponding author}
\address[igm]{LIGM, Universit\'e Paris-Est Marne-la-Vall\'ee,
5, boulevard Descartes, Champs-sur-Marne, 
F-77454 Marne-la-Vall\'ee Cedex 2.}
\address[liafa]{LIAFA, Universit\'e Paris
Diderot--Paris 7 and CNRS, Case 7014, 75205 Paris Cedex 13, France.}
\address[p7]{Universit\'e Paris
Diderot--Paris 7, Case 7014, 75205 Paris Cedex 13, France.}


\begin{abstract}
  In this paper, we prove decidability properties and new results on
  the position of the family of languages generated by (circular)
  splicing systems within the Chomsky hierarchy. The two main results
  of the paper are the following. First, we show that it is decidable,
  given a circular splicing language and a regular language, whether
  they are equal. Second, we prove the language generated by an
  alphabetic splicing system is context-free. Alphabetic splicing
  systems are a generalization of simple and semi-simple splicing
  systems already considered in the literature.
\end{abstract}

\end{frontmatter}


\begin{center}
  \dateandtime
\end{center}
 

\tableofcontents


\setlength\mathindent{2.5\parindent}

\linenumbers

\section{Introduction}

Splicing systems were introduced by T. Head
\cite{Head1987,Head1992,Head1996} as a model of recombination.
The basic operation is to cut words into pieces and to reassemble the
pieces in order to get another word.

There are several variants of splicing systems for circular or linear
words \cite{Head1996}. In this paper, we consider P\u{a}un's circular
splicing, and we introduce a new variant that we call flat splicing.
In both cases, the system is described by an {\em initial set\/} of
words and a {\em finite set of rules\/}. The language generated is the
closure of the initial set under the application of splicing rules.

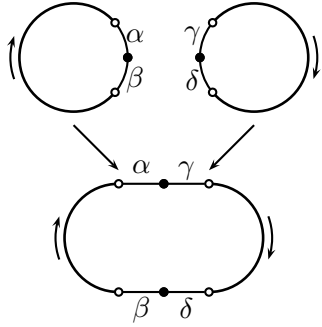
\begin{figure}[htb]
  \centering
\begin{pspicture}(0,-1)(6,6)
\SpecialCoor
\rput(1,5){\psarc[linewidth=1pt](0,0){1.2}{40}{-40}
\psarc{o-*}(0,0){1.2}{-40}{0}
\psarc{*-o}(0,0){1.2}{0}{40}
\psarc{<-}(0,0){1.4}{160}{200}
\rput(1.45;20){$\alpha$}
\rput(1.45;-20){$\beta$}}
\rput(5,5){
\psarc[linewidth=1pt](0,0){1.2}{220}{140}
\psarc{o-*}(0,0){1.2}{140}{180}
\psarc{*-o}(0,0){1.2}{180}{220}
\psarc{<-}(0,0){1.4}{-20}{20}
\rput(1.45;160){$\gamma$}
\rput(1.45;200){$\delta$}}

\rput(2,1){
\psarc[linewidth=1pt](0,0){1.2}{90}{270}
\psline{o-*}(0,1.2)(1.,1.2)
\rput(0.5,1.575){$\alpha$}
\psline{o-*}(0,-1.2)(1.,-1.2)
\rput(0.5,-1.6){$\beta$}
\psarc{<-}(0,0){1.4}{160}{200}
}
\rput(4,1){
\psarc[linewidth=1pt](0,0){1.2}{270}{90}
\psline{o-*}(0,1.2)(-1.,1.2)
\rput(-0.5,1.5){$\gamma$}
\psline{o-*}(0,-1.2)(-1.,-1.2)
\rput(-0.5,-1.55){$\delta$}
\psarc{<-}(0,0){1.4}{-20}{20}
}
\psline{->}(5,3.5)(4,2.5)
\psline{->}(1,3.5)(2,2.5)
\end{pspicture}
  \caption{Circular splicing.} \label{circ}
\end{figure}

\begin{figure}[htb]
  \centering
\begin{pspicture}(-2,0)(10,3)
\rput(0,2){
\psline[linewidth=1pt](-3,0)(-1,0)
\psline[linewidth=1pt](1,0)(3,0)
\psline{o-*}(-1,0)(0,0)
\psline{*-o}(0,0)(1,0)
\rput(-0.5,0.4){$\alpha$}
\rput(0.5,0.4){$\beta$}
}

\rput(8,2){
\psline[linewidth=1pt](-1.5,0)(1.5,0)
\psline{o-*}(1.5,0)(2.5,0)
\psline{*-o}(-2.5,0)(-1.5,0)
\rput(-2,0.4){$\gamma$}
\rput(2,0.4){$\delta$}
}

\rput(1.5,0){
\psline[linewidth=1pt](-3,0)(-1,0)
\psline[linewidth=1pt](6,0)(8,0)
\psline[linewidth=1pt](1,0)(4,0)
\psline{o-*}(-1,0)(0,0)\psline{*-o}(0,0)(1,0)
\psline{o-*}(4,0)(5,0)
\psline{*-o}(5,0)(6,0)
\rput(0.5,0.4){$\gamma$}
\rput(4.5,0.4){$\delta$}
\rput(-0.5,0.4){$\alpha$}
\rput(5.5,0.4){$\beta$}
}

\rput(0,-2){\psline{->}(5.5,3.5)(4.5,2.5)
\psline{->}(2,3.5)(3,2.5)
}
\end{pspicture}
\caption{Flat splicing.}\label{flat}
\end{figure}
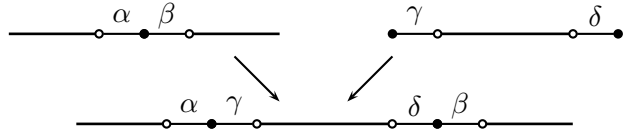

A splicing rule is a quadruplet of words, usually written as
$\Crul\alpha\beta\gamma\delta$. The words $\alpha$, $\beta$, $\gamma$,
$\delta$ are called the {\em handles} of the rule.  A rule indicates
where to cut and what to paste.  More precisely, in a circular
splicing system, given a rule $\Crul{\alpha}{\beta}{\gamma}{\delta}$
and two circular words, the first of the form $u\alpha\cdot\beta v$
and the second of the form $\gamma w \delta$, we cut the first word
between $\alpha$ and $\beta$, the second word between $\delta$ and
$\gamma$ and stick $\alpha$ with $\gamma$ as well as $\delta$ with
$\beta$ in order to get the new circular word $u\alpha \cdot\gamma w
\delta\cdot \beta v$, see Figure \ref{circ}.  The case of flat
splicing systems, which involves linear words, is similar, see Figure
\ref{flat}. In order to emphasize the position where to cut and the
condition on what to paste, we prefer to write
$\rul{\alpha}{\beta}{\gamma}{\delta}$ instead of
$\Crul\alpha\beta\gamma\delta$. This indicates more clearly that one
word is cut between $\alpha$ and $\beta$, and that the word to be
pasted is in $\gamma A^*\delta$.

Our purpose, in introducing flat splicing systems, is to get a direct
approach to standard results in formal language theory. Circular
systems are handled, in a second step, by full linearization.

D. Pixton \cite{Pixton1985,Pixton96} has considered the nature of the
language generated by a splicing system, with some assumptions about
the splicing rules (symmetry, reflexivity and self-splicing). He
proves that the language generated by a splicing system is regular
(resp.  context-free), provided the initial set is regular (resp.
context-free). More generally, if the initial set is in some full AFL,
then the language generated by the system is also in this full AFL.
Without the additional assumptions on the rules, it is known that one
may generate non-regular languages even with a finite initial set
(R. Siromoney, K. G. Subramanian and V. R. Dare \cite{Siromoney92}).
A survey of recent developments along these lines appears
in~\cite{BonizzoniFFZ2010}.

In this paper, we prove decidability properties and new results on the
position of splicing systems and their languages within the Chomsky
hierarchy.  We introduce a special class of splicing rules called
alphabetic rules. A rule is \emph{alphabetic} if its four handles are
letters or the empty word. A splicing system is alphabetic when all
its rules are alphabetic.  Special cases of alphabetic splicing
systems, called \emph{simple} or \emph{semi-simple} systems, have been
considered in the literature
\cite{CeterchiMS2004,Ceterchi2006,BonizzoniFZ2009}. In a 
semi-simple system, all rules $\Crul\alpha\beta\gamma\delta$ satisfy
the condition that the words $\alpha\beta$ and $\gamma\delta$ are
letters. In a simple system, one requires in addition that these
letters are equal, that is $\alpha\beta=\gamma\delta$.

We show that alphabetic systems have several
remarkable properties that do not hold for general systems.

We consider first the problem of deciding whether the language
generated by a splicing system is regular. The problem is still open,
but has been solved in special cases \cite{Bonizzoni2010}. Our
contribution is the following (Theorem~\ref{theodecidable}). It is
decidable, given a circular splicing language and a regular language,
whether they are equal. The corresponding inclusion problems are still
open.  We also show (Remark~\ref{rem-decidable}) that it is decidable
whether a given regular language is an alphabetic splicing
language. This is related to another problem that we do not consider
here, namely to give a characterization of those regular languages
that are splicing languages, or vice-versa. For recent results 
see~\cite{Bonizzoni2010}, and for a survey see~\cite{BonizzoniFFZ2010}.

The next problem we consider concerns the comparison of the family of
splicing languages with the Chomsky hierarchy. We first prove
(Theorem~\ref{theo-contextsensitive}) that splicing languages are
always context-sensitive. Next, we prove, and this is the main result
of the paper (Theorem~\ref{context-fre}), that alphabetic splicing
languages are context-free.  The proof of this result is in several steps.

We consider first a special class of systems called \emph{pure}, and
we prove (Theorem~\ref{Insertion}) that pure alphabetic systems
generate context-free languages, even if the initial set is itself
context-free.

We next consider another special class of systems called
\emph{concatenation} systems. In those systems, insertions always
take place at one end of the word. We show
(Theorem~\ref{Concatenation}) the language generated by a
concatenation system is context-free, even if the initial set is itself
context-free.

The next step is to mix these two kinds of splicing systems. We call
them \emph{heterogeneous} systems. Every alphabetic splicing system is
heterogeneous. The key observation, for the proof of the main result,
is that in a heterogeneous system, all concatenations can be executed
before any proper insertion (Lemma~\ref{exchange}). We call this a
weak commutation property. The main theorem then easily follows.

The relation between circular and flat alphabetic splicing systems is
described in Proposition~\ref{propCirc}. It follows easily that the
main result also holds for circular splicing (Theorem~\ref{theo-circularContext-free}).

The proofs rely on
so-called generalized context-free grammars, a notion that is rather
old but seems not to be well-known. All proofs are effective.

The paper is organized as follows. We start by introducing the new type
of splicing systems called flat splicing systems: these systems behave
like circular systems, but operate on linear words. 

In Section~3, we prove the
decidability result mentioned earlier
(Theorem~\ref{theodecidable}). Section~\ref{sec:cs} contains the
proof that splicing languages are always context-sensitive.

Section~\ref{sec:alphabetic} defines alphabetic splicing systems and
states the main result (Theorem~\ref{context-fre}) namely that the
language generated by a flat or circular  alphabetic splicing system is context-free,
even if the initial set is context-free. In this section, a
normalization of splicing systems called completion is presented. The
complete systems defined here are not the same as the complete systems
in~\cite{BonizzoniFZ2009}. 

Section~\ref{sec-pure} introduces pure splicing systems. Here, it is
proved that the language generated by a context-free pure splicing
system is context-free.  The proof uses some results on context-free
languages which are recalled in Section~\ref{subsec-contextfree}.

In the next section (Section~\ref{sec-echange}), we first define
concatenation systems and prove that (alphabetic) concatenation
systems produce only context-free languages. Then heterogeneous
systems are defined, and the weak commutation lemma
(Lemma~\ref{exchange}) is proved. This section ends with the proof of
the main theorem for flat splicing systems.

Section~\ref{circular} describes the relationship between flat and
circular splicing systems and their languages. It contains the 
proof of the main theorem for circular splicing systems.

The proofs that context-free alphabetic concatenation and pure systems
generate context-free languages is done by giving explicitly the
grammars. These grammars deviate from the standard form of grammars by
the fact that the set of derivation rules may be infinite, provided they
are themselves context-free sets. It is an old result from formal
language theory \cite{Kral70} that generalized context-free grammars
of this kind still generated context-free languages. For sake of
completeness, we include a sketch of the proof of this result,
together with an example, in an appendix.

The results of the paper were announced by the third author in
\cite{Fagnot2004}. 

\section{Definitions}
\label{section:def}
\subsection{Words, circular words}

As usual, an {\em alphabet} $A$ is a finite set of {\em letters}.  A
word $u= u_0u_1\ldots u_{n-1}$ is a finite sequence of letters. When
it is useful to compare words with circular words defined below, they
will be called {\em linear} words.

Two words $u$ and $v$ are conjugate, denoted by $u \sim v$ if there
exist two words $x$ and $y$ such that $u=xy$ and $v=yx$. This is an
equivalence relation. A {\em circular word} is an equivalence class of
$\sim$, that is an element of the quotient of $A^*$ by the
relation~$\sim$.  The equivalence class of $u$ will be noted
$\cir{u}$.  We also say that $\cir{u}$ is the {\em circularization} of
$u$.  For example, $\cir{abb} = \{abb, bab,bba\}$ is a circular word.
A circular word can be viewed as in Figure \ref{circ} as a word
written on a circle.  A set of circular words is a \emph{circular
  language}.

Let $C$ be a language of circular words.  Its {\em full
  linearization}, denoted by $\lin{C}$, is the language
 $\lin{C} = \{ u \in A^*\mid \cir{u} \in C\}$.

Let $L$ be a language of linear words. Its {\em circularization}
$\cir{L}$ is equal to $\cir{L} =\{\cir u\mid u\in L\}$.

A language $L$ of circular words is {\em regular} (resp. {\em context-free}, resp. 
{\em context-sensitive}) if  its full linearization is  regular (resp.  context-free, 
resp.  context-sensitive).

Let $G$ be a grammar, the language generated by $G$ will be denoted by
$\lang{G}$.  Let $S$ be a non-terminal symbol, we will denote
$\sublang{G}{S}$ the language produced by the grammar $G$ with $S$ as
axiom.

\subsection{Splicing systems}

We start with a short description of circular splicing systems. These
systems are well known, see e.g. \cite{Bonizzoni2010}.  Then we present flat
splicing systems which are new systems. They are of interest for
proving language-theoretic results because they allow us to separate
operations on formal languages and grammars from the operation of
circular closure (circularization).  It appears that proofs for linear
words are sometimes simpler because they rely directly on standard
background on formal languages.

\subsubsection{Circular splicing systems}

A circular splicing system is a triplet $\cal S=(A,I, R)$, where $A$
is an alphabet, $I$ is a set of circular words on $A$, called {\em
  initial set} and $R$ is a finite set of {\em splicing rules}, which
are quadruplets $\rul{\alpha}{\beta}{\gamma}{\delta}$ of {\em linear}
words on $A$. The words $\alpha, \beta, \gamma$ and $\delta$ are
called the {\em handles} of the rule. In the literature (see
e.g.~\cite{Bonizzoni2010}), a rule is written as
$\alpha\#\beta\$\gamma\#\delta$.

If $r = \rul{\alpha}{\beta}{\gamma}{\delta}$ is a splicing rule then the circular words
$\cir u =\cir{(\beta x\alpha)}$ and $\cir v = \cir{(\gamma y\delta)}$  
produce the circular word
$\cir w = \cir{(\beta x \alpha\gamma y \delta)}$.
We will denote this production by $\produc{r}{\cir u}{\cir v}{\cir w}$.
The \emph{language generated} by the circular splicing system is the
smallest language $C$ of circular words containing $I$ and closed by
$R$, i.e., such that for any couple of words $\cir u$ and $\cir v$ in
$C$ and any rule $r$ in $R$, any circular word $\cir w$ such that
$\produc{r}{\cir u}{\cir v}{\cir w}$ is also in $C$. This set of
circular words is denoted by $\langc{{\cal S}}$.

A circular splicing system is \emph{finite} (resp. \emph{regular,
  context-free, context-sensitive}) if its initial set is finite
(resp. regular, context-free, context-sensitive).

A splicing rule $r = \rul{\alpha}{\beta}{\gamma}{\delta}$ is
\emph{alphabetic} if its four handles $\alpha, \beta, \gamma$ and
$\delta$ are letters or the empty word. A circular splicing system is
alphabetic if all its rules are alphabetic.

\begin{example} \label{sir_ex2} Let ${\cal S}=(A,I,R)$ be the (finite
  alphabetic) circular splicing system defined by $I= \{\cir{(ab)}\}$
  and $R = \{ \rul{a}{b}{a}{b} \}$. It produces the context-free
  language $\langc{{\cal S'}}= \{ \cir{(a^nb^n)} \mid n \geq 1\}$.
\end{example}

\subsubsection{Flat splicing systems}

A \emph{flat splicing system}, or a \emph{splicing system} for short, is a
triplet $\cal S=(A,I, R)$, where $A$ is an alphabet, $I$ is a set of
words over $A$, called the {\em initial set} and $R$ is a finite set
of {\em splicing rules}, which are quadruplets
$\rul{\alpha}{\beta}{\gamma}{\delta}$ of words over $A$.  Again, a
rule is \emph{alphabetic} if its the four handles $\alpha, \beta,
\gamma$ and $\delta$ are letters or the empty word. A splicing system
is alphabetic if all its rules are alphabetic.

Let $r = \rul{\alpha}{\beta}{\gamma}{\delta}$ be a splicing
rule. Given two words $u = x \alpha\cdot\beta y$ and $v = \gamma z
\delta$, applying $r$ to the pair $(u,v)$ yields the word $w= x
\alpha\cdot\gamma z\delta\cdot \beta y $. (The dots are used only to
mark the places of cutting and pasting, they are not parts of the
words.)  This operation is denoted by $\produc{r}{u}{v}{w}$ and is
called a {\em production}. Note that the first word (here $u$) is
always the one in which the second word (here $v$) is inserted.

\begin{example}
1.  Consider the splicing rule $r=\rul{ab}{c}{aa}{b}$. We have the
  production $\produc{r}{bab \cdot cc}{aaccb}{bab \cdot aaccb \cdot cc
  }$.  

2. Consider the splicing rule $\rul{b}{b}{a}{a}$. Note that we  cannot
produce the word 
$b\cdot a \cdot b$ from the word $b \cdot b$ and the singleton $a$, 
because the rule requires that the inserted word has at least two
letters. On the contrary, the rule
$\rul{b}{b}{\e}{a}$ does produce the word $bab$ from the
words $bb$ and $a$. 

3. For the rule $r=\rul{\e}{b}{a}{a}$, the production 
 $\produc{r}{\cdot bbc}{aba}{aba\cdot bbc}$,
is in fact a concatenation.

4. As a final example, the rule   $\rul{\e}{\e}{\e}{\e}$ 
permits all insertions of a word into another one.
\end{example}

The \emph{language generated} by the flat splicing system ${\cal S}
=(A,I, R)$, denoted $\langf{{\cal S}}$, is the smallest language $L$
containing $I$ and closed by $R$, i.e., such that for any couple of
words $u$ and $v$ in $L$ and any rule $r$ in $R$, then any word such
that $\produc{r}{u}{v}{w}$ is also in $L$.

\begin{example} \label{sir_ex} Consider the splicing system over $A =
  \{a, b\}$ with initial set $I= \{ab\}$ and the unique splicing rule
  $r =\rul{a}{b}{a}{b}$.  It generates the context-free and
  non-regular language $\langf{{\cal S}}= \{ a^nb^n \mid n \geq 1\}$.
\end{example}

\begin{rema}\label{epsilon}
  A production $\produc{r}{u}{v}{w}$, where $u=\e$ or $v=\e$, even
  when it is permitted, is
  useless. Indeed, one has $\produc{r}{\e}{v}{v}$ and
  $\produc{r}{u}{\e}{u}$. As a consequence, given a splicing system
  $\cal S=(A,I, R)$ one has $\e\in\langf{\cal S}$ if and only
  $\e\in I$. So we can assume that $\e\notin I$ without loss of
  generality.  This remark holds also for circular splicing systems.
\end{rema}

\begin{rema}\label{singleton}
  A production $\produc{r}{u}{v}{w}$, where $|w|=1$, even when it is
  permitted, is useless. Indeed, since $|u|+|v|=|w|$, one has in this
  case $w=u$ or $w=v$.  As a consequence, given a splicing system
  $\cal S=(A,I, R)$, and a letter $a\in A$, one has $a\in\langf{\cal
    S}$ if and only $a\in I$. However, we cannot assume that $a\notin
  I$ without possibly changing the language it generates.  This
  remark holds also for circular splicing systems.
\end{rema}

\begin{rema}
  Flat splicing is different from linear splicing as it is defined in
  \cite{Pixton1985}.
\end{rema}

\begin{rema}
  Let ${\cal S} =(A,I, R) $ be a flat splicing system and let ${\cal
    S'} =(A,\cir{I},R) $ be the circular splicing system with the same
  splicing rules.  The full linearization of $\langc{{\cal S'}}$ is
  the closure of the linear language $I$ under the composition of the
  two operations of circularization and splicing. However, it does not
  suffice, in general, to just consider a single
  circularization. Indeed, the equality $\langc{{\cal S'}} =
  \cir{\langf{{\cal S}}}$ does not hold in general. However, the inclusion $
  \cir{\langf{{\cal S}}} \subseteq \langc{{\cal S'}}$ is always true.

  Consider the flat splicing system over $A =\{a,b\}$, initial set $I
  = \{ba\}$ and with the single rule $\rul abab$.  Clearly, the rule
  cannot be applied, and consequently the language generated by the
  system reduces to $I$, and its circularization gives $\cir I$. The
  circular language generated by the system is $\cir\{a^nb^n\mid
  n\ge1\}$, which is much larger than $\cir I$.
\end{rema}


\section{A decision problem}


In this section, we prove the following result.

\begin{theo}\label{theodecidable}
  Given a regular circular (resp. flat) splicing system ${\cal S}$ and
  a regular language $K$, it is decidable whether $\langc{{\cal S}} =
  K$ (resp. $\langf{{\cal S}} = K$).
\end{theo}

\begin{preuve}
  We assume that neither $I$ nor $K$ contains $\e$, since otherwise it
  suffices, according to Remark \ref{epsilon}, to check that $\e$ is
  contained in both sets.

  Let ${\cal S} =(A,I,R)$.  Let ${\cal A} = (A, Q, q_o, Q_F)$ be a
  deterministic automaton recognizing $K$, with $Q$ the set of states,
  $q_o$ the initial state and $Q_F$ the set of final states.  The
  transition function is denoted by ``$\cdot$'' in the following way:
  for a state $q$ and a word $v$, $q \cdot v$ denotes the state that
  is reached by $v$ from $q$.

  For any state $q \in Q$, we define $G_{q}=\{v \mid q_o \cdot v =
  q\}$ and $D_{q}=\{v \mid q \cdot v \in Q_F\}$.  The set $G_q$ is the
  set of all words which label paths from $q_0$ to $q$, and $D_{q}$ is
  the set of all words which label paths from $q$ to a terminal state.
  Both sets are regular.

  Next, let $P = \{w \in A^*\mid \produc{r}{u}{v}{w}, r\in R, u,v \in
  K \}$. The set $P$ is the set of the words that can be obtained by
  splicing two words of $K$.

For each rule $r= \rul{\alpha}{\beta}{\gamma}{\delta} $,
let
\begin{align*}
  K_r = &
  \{w \in A^* \mid \produc{r}{u}{v}{w}\,, \mbox{ with } u,v \in K\} \\
  = &\{ x \alpha\cdot\gamma z\delta\cdot\beta y \mid x
  \alpha\cdot\beta y \in K, \gamma z \delta \in K,\, x,y,z \in
  A^*\}\,.
\end{align*}
It is easily checked that
\begin{displaymath}
  K_r =\bigcup_{q\in Q} (G_q \cap A^*\alpha) (K \cap \gamma
  A^* \delta)  (D_q \cap \beta A^* )\,. 
\end{displaymath}
This expression shows that each language $K_r$ is regular, and so is
$P = \bigcup_{r\in R}K_r$ because $R$ is finite.

We first consider flat splicing system.  The algorithm consists in
checking three inclusions. We claim that $\langf{{\cal S}} = K$ if and
only if the following three inclusions hold.
\begin{enumerate}
\item[(1)] $I \subseteq K$,
\item[(2)] $P \subseteq K$,
\item[(3)] $K \setminus P\subseteq I$.
\end{enumerate}
Take the claim for granted. Then the equality $\langf{{\cal S}} = K$ is
decidable since the three inclusions, that involve only regular
languages, are decidable.

Now we prove the claim, namely that $\langf{{\cal S}} = K$ if and
only if the above-mentioned three inclusions hold.

If $\langf{{\cal S}} = K$, then  (1), (2) and (3)  are obviously true.

Conversely, assume now that these three inclusions hold.
Since   $I \subseteq K$ by (1) and since $K$ is closed under the rules
of splicing of $R$ by (2),  obviously $\langf{{\cal S}} \subseteq K$.

Next, we prove the reverse inclusion $ K  \subseteq \langf{{\cal S}}$
by induction on the length of the words in $K$. Let $w\in K$. Since
$P\subset K$ by~(2), one has $K=P\cup (K\setminus P)$.
If $w\in K\setminus P$, then
by~(3), $w\in I$ and therefore $w\in \langf{{\cal S}}$.
Otherwise, there are words $u,v\in K$ of shorter length such that
$\produc{r}{u}{v}{w}$ for some $r\in R$. By induction, $u,v\in
\langf{{\cal S}}$
and consequently 
$w\in \langf{{\cal S}}$.

For circular splicing systems, it suffices to check, in addition, that
$K$ is closed under conjugacy and to replace $K_r$ by $\cir{(K_r)}$.
\end{preuve}

\begin{rema}
  There are two related problems which are still open. The first is to
  decide whether the language generated by a splicing system is
  regular, and the second is to decide whether a regular language can
  be generated by a splicing system. We shall see below that the second
  problem is decidable in the case of what we call alphabetic splicing systems.
\end{rema}

\begin{rema}
  The inclusion problems, for both inclusions, i.e., the problem of
  deciding whether $\langf{{\cal S}} \subseteq K$ or whether $
  K\subseteq\langf{{\cal S}}$ (resp. $\langc{{\cal S}} \subseteq K$ or
  $ K\subseteq\langc{{\cal S}}$) are still open.
\end{rema}

\begin{rema}
  The characterization of the family of regular languages which can be
  obtained by a circular splicing system, is still open.  However,
  partial results have been obtained by P. Bonizzoni, C. De Felice,
  G. Mauri and R. Zizza \cite{Bonizzoni2000,Bonizzoni2002,Bonizzoni2010}.  In
  particular, a complete characterization of 
  languages over one letter
generated by a splicing system is given in
  \cite{Bonizzoni2002}.  Recently, a description of the languages
  generated by a family of alphabetic splicing systems called
  semi-simple systems has been given in~\cite{Bonizzoni2010}.
\end{rema}

\begin{rema}\label{rem-decidable}
  Given a regular language $K$ over an alphabet $A$, it is decidable
  whether it can be generated by a finite alphabetic splicing
  system. (The problem is meaningless for regular systems.) Indeed,
  observe first that there are only finitely many alphabetic splicing
  rules over $A$. So there are only finitely many sets of alphabetic
  splicing rules over $A$. Choose one such set and call it $R$. Define
  $P$ as in the proof of Theorem~\ref{theodecidable}. If $P\not\subset
  K$ or $K\setminus P$ is infinite, then the test is
  negative. Otherwise, the splicing system ${\cal S}=(A,K\setminus P,
  R)$ generates $K$.
\end{rema}


\section{Splicing languages are  context-sensitive}\label{sec:cs}

We will see that the highest level in Chomsky hierarchy which can be
obtained by splicing systems with a finite 
initial set and a finite set of rules is the context-sensitive level.
This result remains true when  the initial set is context-sensitive.
 
Before proving this property, we give an example 
of a splicing language which is not context-free.

\subsection{A splicing language which is not context-free}

We first consider flat splicing.

Let  $A$ be the alphabet $ \{0,1,2,3, \bb , \ff  \}$ and set $u=0123$.
Let ${\cal S}= (A,I,R)$ be the flat splicing system
with
\begin{displaymath}
I = \{ \bb u\ff , 0, 1, 2, 3 \}
\end{displaymath}
and with $R$ composed of the rules
\begin{align*}
  &\rul{\bb }{u}{0}{\e}, &\hspace{-2cm}&\rul{0u}{u}{0}{\e},  \\
  &\rul{0}{u\ff }{1}{\e},&\hspace{-2cm}& \rul{0}{u01u}{1}{\e}, \\ 
  &\rul{\bb 01 }{u}{2}{\e}, &\hspace{-2cm}&\rul{012u01}{u }{2}{\e} ,\\
  &\rul{012 }{u\ff }{3}{\e},&\hspace{-2cm}&\rul{012}{uuu}{3}{\e}\,.
 \end{align*}
 This splicing system produces the language
\begin{align*}
  \langf{{\cal S}}  = &\{ \bb  (u)^{2^n}\ff   \mid n \geq 0\}\\
  & \cup \{ \bb  (0u)^p(u)^q\ff   \mid p+q = 2^n, n \geq 0\}\\
  & \cup \{ \bb  (0u)^p(01u)^q\ff   \mid p+q = 2^n, n \geq 0\} \\
  & \cup \{ \bb  (012u)^p(01u)^q\ff   \mid p+q = 2^n, n \geq 0\}\\
  &\cup\{ \bb (012u)^p(uu)^q\ff \mid p+q = 2^n, n \geq 0\}\,.
 \end{align*}
Indeed, given a word $\bb u^n\ff$, the first two rules of $R$ generate
a left-to-right sweep inserting the symbol $0$ in head of each $u$:
\begin{displaymath}
  \bb u^n\ff\to\bb(0u)u^{n-1}\ff\to\cdots\to\bb(0u)^{n-1}u\ff\to\bb(0u)^n\ff\,.
\end{displaymath}
(We write here $x\to y$ instead of $\produc {}x{0}y$.) The next two
rules generate a right-to-left sweep which inserts a symbol $1$
in head of each~$u$. This gives
\begin{displaymath}
  \bb (0u)^n\ff\to\bb(01u)(0u)^{n-1}\ff\to\cdots\to\bb(01u)^{n-1}0u\ff\to\bb(01u)^n\ff\,.
\end{displaymath}
The next two rules are used to insert a symbol $2$ in head of each
$u$, again in a left-to-right sweep. This gives the word $\bb
(012u)^n\ff$. Finally, the last two rules insert a $3$ in head of each
$u$. The final result is $\bb u^{2n}\ff$.

The intersection of the language $\langf{{\cal S}}$ with the regular
language $\bb (u)^*\ff $ is equal to $\{ \bb (u)^{2^n}\ff \mid n \geq
1\}$. The latter language is not context-free.

Concerning circular splicing systems, recall that a circular language
is con\-text-sensitive if and only if its full linearization is
context-sensitive.  If we take the circular splicing system with the
same rules and the same initial language, we can check that the
language $\langc{{\cal S}}$ is such that $\langc{{\cal S}} \cap \bb
(u)^*\ff =\langf{{\cal S}} $.  Thus, we also can produce a language
which is not context-free  with a circular splicing system.

\subsection{Splicing  languages are always context-sensitive}

\begin{theo}\label{theo-contextsensitive}
  The language generated by a context-sensitive circular (resp. flat)
  splicing system is context-sensitive.
\end{theo}

The proof uses bounded automata.  Recall that a $k$-\emph{linear
  bounded automaton} ($k$-LBA) is a non-deterministic Turing machine
with a tape of only $kn$ cells, where $n$ is the size of the input.
We will use in the sequel the following characterization of
context-sensitive languages.  A language is context-sensitive if and
only if it is recognized by a $k$-LBA (see, for example,
\cite{Harrison78}).  It is known that it is always possible to
recognize a context-sensitive language with a $1$-LBA.\medskip

\begin{preuve} 
We start with the case of a flat system.
Let ${\cal S}= (A,I,R)$ be a flat splicing system.
Let $\cal T$ a $1$-LBA recognizing $I$.
We construct a $3$-LBA machine $\cal U$ which  recognizes the language $\langf{{\cal S}}$.

Let $u$ be the word written on the tape at the beginning of the
computation. Let $\#$ be a new symbol.
The machine  works as follows.

During the computation the  word written on the tape has the form
\begin{displaymath}
u_1 \# u_2 \# \cdots  \# u_{n-1} \# u_n\,,
\end{displaymath}
where the $u_i$ are words on the alphabet $A$.

Repeat the following operation as long as possible.
\begin{enumerate}
\item[(1)] If the tape is void, stop and return
  ``yes''.
\item[(2)] If $u_n$ is in the set $I$ (this test is performed by machine
  $\cal T$), remove $u_n$ along with the symbol $\#$ which may precede
  $u_n$.
\item[(3)] Choose randomly a rule $r =
  \rul{\alpha}{\beta}{\gamma}{\delta}$ in $R$, and choose randomly, if
  it exists, a decomposition of $u_n$ of the form $u_n = x\alpha
  \gamma y \delta \beta z$ such that neither $x\alpha \beta z$ nor
  $\gamma y \delta$ are empty word.  Remove the subword $\gamma y
  \delta$ from $u_n$ and place it at the right after a $\#$
  symbol. Then shift the string $ \beta z \# \gamma y \delta$ so that
  we have on the tape $u_1\# u_2 \# \cdots \# u_{n-1} \# x\alpha \beta
  z \# \gamma y \delta$. If no choice exists, stop the computation.
\end{enumerate}
It can be easily seen that the length of the tape is always less that
$3 |u|$.
If no computation succeeds, then the word is rejected.

In the case of a circular splicing system, the method is almost the
same. The only difference is that, in the last step, one chooses in
addition randomly one of the conjugates of $u_n$.
\end{preuve}


\section{Alphabetic splicing systems}\label{sec:alphabetic}


A rule in a splicing system is called {\em alphabetic} if its handles
have length at most one. A splicing system is called {\em alphabetic}
if all its rules are alphabetic.

The splicing systems of Examples~\ref{sir_ex2} and~\ref{sir_ex} are
alphabetic. They generate a non-regular language, although they have a finite 
initial set. Let us give another example.

\begin{example} \label{dyck_ex} Let ${\cal S}=(A,I,R)$ be the flat
  splicing system defined by $A = \{a,\bar{a} \}$, $I= \{a\bar{a}\}$
  and $R = \{ \rul{\e}{\e}{\e}
  {\e} \}$. It generates the Dyck language.  Recall that the
  Dyck language over $\{a, \bar{a}\}$ is the language of parenthesized
  expressions, $a, \bar{a}$ being viewed as a pair of matching
  parentheses.

  The circular splicing system ${\cal S}=(A,I,R)$ defined by $A =
  \{a,\bar{a} \}$, $I= \{\cir{(a\bar{a})}\}$ and $R = \{
  \rul{\e}{\e}{\e}{\e} \}$
  generates the language $\hat{D}$ of words having as many $a$ as
  $\bar{a}$.  The language $\hat{D}$ is the circularization of the
  Dyck language.
\end{example}

\begin{rema}
  All examples given so far show that alphabetic splicing systems
  generate always a context-free languages, and this is indeed the
  main result of the paper. Observe however that we cannot get all
  context-free languages as splicing languages with a finite initial
  set. For example, the language $L =\{ a^nb^nc \mid n \geq n\}$
  cannot be obtained by such a splicing system. (Consider indeed the
  fact that all words in $L$ have the same number of $c$.)
\end{rema}

\subsection{Main theorem}

We now state the main theorem, namely that alphabetic rules and a
context-free initial set can produce only context-free languages.

\begin{theo} \label{context-fre} 
  {\upshape{(i)}} The language generated by a circular alphabetic
  context-free splicing system is context-free.\\
  {\upshape{(ii)}} The language generated by a flat alphabetic
  context-free splicing system is context-free.
\end{theo}

This theorem is effective, that is, we can actually construct a
context-free grammar which generates the language produced by the
splicing system.
The rest of the paper is devoted to the  proof of this theorem. 

Section~\ref{sec-pure} introduces pure splicing systems. Here, it is
proved that the language generated by a context-free pure splicing
system is context-free.  The proof uses some results on context-free
languages which are recalled in Section~\ref{subsec-contextfree}.

In the next section (Section~\ref{sec-echange}), we first define
concatenation systems and prove that (alphabetic) concatenation
systems produce only context-free languages. Then heterogeneous
systems are defined, and the weak commutation lemma
(Lemma~\ref{exchange}) is proved. This section ends with the proof of
the main theorem for flat splicing systems.

Section~\ref{circular} describes the relationship between flat and
circular splicing systems and their languages. It contains the 
proof of the main theorem for circular splicing systems.

The proofs that concatenation systems and alphabetic pure
systems generate con\-text-free languages are done by giving explicitly the
grammars. These grammars deviate from the standard form of grammars by
the fact that the sets of derivation
 rules may be infinite, provided they
are themselves context-free sets. It is an old result from formal
language theory \cite{Kral70} that generalized context-free grammars
of this kind still generate context-free languages. For the sake of
completeness, we include a sketch of the proof of this result,
together with an example, in an appendix.

We start with a technical normalization of splicing systems.

\subsection{Complete set of rules}

Completion of rules is a tool to manage the usage of the empty word
$\e$ among the handles $\alpha,\beta,\gamma,\delta$ of an alphabetic  rule
\begin{displaymath}
  r = \rul{\alpha}{\beta}{\gamma}{\delta}
\end{displaymath}
in a production
\begin{equation}\label{prod}
  \produc r u v w \,.
\end{equation}
Assume first that $\delta=\e$. (The case where $\gamma=\e$ is
symmetric.) In this case, the production~\eqref{prod}  is valid provided
$v$ starts with
$\gamma$ (and of course if $u$ has an appropriate factorization
$u=x\alpha\beta y$). Let $d$ be the final letter of $v$.
Then the same result is obtained with the rule
\begin{displaymath}
  r_d= \rul\alpha\beta\gamma d\,,
\end{displaymath}
with only one, but noticeable exception: this is the case where $v$ is
a single letter, that is $v=\gamma$. Observe that this may happen only
if $v$ is in the initial set of the system.

In other words, a production
\begin{displaymath}
    r = \rul{\alpha}{\beta}{\gamma}\e
\end{displaymath}
is mandatory if and only if $\gamma\in I$. For all words $v\ne\gamma$,
the production~\eqref{prod}  is realized by the use of the rule $r_d$
where $d$ is the final letter of $v$. Thus a rule with $\delta=\e$ can
be replaced by the set of rules $r_d$, for $d\in A$ with one
exception.

Assume next that $\beta=\e$. (The case where $\alpha=\e$ is symmetric.)
In this case, the production
\begin{displaymath}
    \produc r u v w
\end{displaymath}
is valid provided $\alpha$ occurs in $u$ (and $v$ begins with $\gamma$
and ends with $\delta$). This holds in particular when $\alpha$ is the
final letter of $u$. In this case, one gets
\begin{displaymath}
  w=uv\,.
\end{displaymath}
In other words, the application of the rule reduces to a simple
concatenation. If however $u$ has another occurrence of $\alpha$, that
is if $u=x\alpha y$ for some $y\ne\e$, then the rule $r$ can be
replaced by the appropriate rule $r_d=\rul\alpha d\gamma\delta$, where
$d$ is the initial letter of $y$.

In conclusion, the use of a rule
\begin{displaymath}
   r = \rul{\alpha}{\beta}{\gamma}\e \quad(\text{resp. }
   r = \rul{\alpha}{\beta}{\e}\delta)
\end{displaymath}
can always be replaced by the use of a rule 
\begin{displaymath}
     r = \rul{\alpha}{\beta}{\gamma} d \quad(\text{resp. }
   r = \rul{\alpha}{\beta}c\delta)
\end{displaymath}
for letters $c,d\in A$, except -- and this is the only case -- when
the word to be inserted is a single letter which is in the initial set.

On the contrary, the use of a rule
\begin{displaymath}
   r = \rul{\alpha}{\e}{\gamma}\delta \quad(\text{resp. }
   r = \rul{\e}{\beta}{\gamma}\delta)
\end{displaymath}
can be replaced by the use of a rule
\begin{displaymath}
   r = \rul{\alpha}{b}{\gamma}\delta \quad(\text{resp. }
   r = \rul{a}{\beta}{\gamma}\delta)
\end{displaymath}
for letters $a,b\in A$, except when the result is a concatenation
$w=uv$ (resp. $w=vu$).

\begin{example}
  Let ${\cal S}=(A,I,R)$ with $A = \{a,b, c\}$, $I =\{abc, abb\}$ and
  $R$ composed of the single rule $r = \rul{b}{\e}{a}{b}$.  The rule
  $r$ permits the production $\produc{r}{ab\cdot c}{abb}{ab\cdot
    abb\cdot c}$. This production could also be realized with the rule
  $r' = \rul{b}{c}{a}{b}$ obtained from $r$ by replacing $\e$ by $c$.
  Similarly, the production $\produc{r}{ab\cdot b}{abb}{ab\cdot
    abb\cdot b}$ could also be realized with the rule $r'' =
  \rul{b}{b}{a}{b}$.  Conversely, all productions that can be realized
  with $r'$ and $r''$ can be made with $r$.

  We can thus check that the system ${\cal S'}=(A,I,R')$ with the set
  of rules $R' = \{ \rul{b}{\e}{a}{b}, \rul{b}{a}{a}{b},
  \rul{b}{b}{a}{b} , \rul{b}{c}{a}{b}\}$ produces the same language as
  the system ${\cal S}$ does.

  However, the production $\produc{r}{abb\cdot }{abb}{abb \cdot abb}$,
  cannot be obtained by use of a production without $\e$-handle.  So,
  the system ${\cal S''}=(A,I,R'')$ with $R'' = \{ \rul{b}{a}{a}{b},
  \rul{b}{b}{a}{b} , \rul{b}{c}{a}{b}\}$ does not produce the same
  language as the system ${\cal S}$ does.
\end{example}

We say that a splicing system ${\cal S}=(A,I,R)$ is {\em complete} if
for any rule $r= \rul{\alpha_1}{\alpha_2}{\alpha_3}{\alpha_4}$ in $R$,
whenever one or several of the $\alpha_i$ are equal to the empty word,
then the set $R$ contains all rules obtained by replacing some or all
of the empty handles by all letters of the alphabet.

For example, the system  ${\cal S}=(A,I,R')$ is complete.
The \emph{completion} of a splicing system consists in adding to the
system the rules that makes it complete. Completion is possible for
alphabetic splicing systems without changing the language it produces.

\begin{lem}
  For any alphabetic splicing system ${\cal S}=(A,I,R)$, the complete
  alphabetic splicing system ${\cal \hat{S}}=(A,I,\hat R)$ obtained by
  completing the set of productions generates the same language.
\end{lem}
The proof is left to the reader.

Observe that completion may increase considerably the number of rules
of a splicing system. Thus, over a $k$-letter alphabet, completing a
rule with one $\e$-handle adds $k$ rules, and if the rule has two
$\e$-handles, completion adds $k^2+2k$ rules\dots

In the proof of Theorem \ref{context-fre}, i.e., in Sections
\ref{sec-echange}, 
\ref{circular} we will assume that splicing systems are
complete.

\begin{rema}
  Complete systems may simplify some verifications.  Thus, in order
  to verify that one may insert a letter $a$ between some letters~$d$
  and~$b$, it suffices to check that one of $\rul{d}{b}{a}{\e}$ or
  $\rul{d}{b}{\e}{a}$ is in the set of splicing rules.  Otherwise we
  would also have to check whether one of $\rul{\e}{b}{a}{\e}$ or
  $\rul{\e}{\e}{a}{\e}$ or $\rul{d}{b}{\e}{\e}$,\dots is in the set
  of splicing rules.
\end{rema}

\section{Pure splicing systems\label{sec-pure}}

In this section, we consider a subclass of splicing systems called
pure systems, and we prove (Theorem~\ref{Insertion}) that these
systems generate context-free languages. We start with a description
of two theorems for context-free languages that will be useful.

\subsection{Two  theorems on context-free languages}\label{subsec-contextfree}

We recall here, for the convenience of the reader, the notion of
context-free substitutions, generalized context-free grammars along
with two substitution theorems.  A sketch of proof of the second
theorem and an example are given in the appendix.

Let $A$ and $B$ be two alphabets.  A {\em substitution} from $A^*$ to
$B^*$ is a mapping $\sigma$ from m $A^*$ into subsets of $B^*$ such
that $\sigma(\e) = \{\e\}$ and
\begin{displaymath}
  \sigma(xy) = \sigma(x)\sigma(y)
\end{displaymath}
for all $x,y \in A^*$. The product of the right-hand side
is the product of subsets of $B^*$.  The substitution is called {\em
  finite} (resp. {\em regular, context-free, context-sensitive}) if
all the languages $\sigma(a)$, for $a$ letter of $A$, are {\em finite}
(resp. {\em regular, context-free, context-sensitive}).

The usual substitution theorem for context-free languages (see, for
example, \cite{Harrison78}) is the following.
\begin{theo} \label{subst1} Let $L$ be a context-free language over an
  alphabet $A$ and let $\sigma$ be a context-free substitution. Then the
  language $\sigma(L)$ is context-free.
\end{theo}

A more general theorem, which is also a kind of substitution theorem,
is due to J. Kr\`al \cite{Kral70}.  In order to state it, we introduce
the following definition.  A {\em generalized grammar} $G$ is a
quadruplet $(A, V, S, R)$, where $A$ is a terminal alphabet, $V$ is a
non-terminal alphabet, and $S\in V$ is the axiom. The set of rules $R$
is a possibly infinite subset of $V\times (A \cup V)^*$.  For each,
$v\in V$, define $M_v = \{ m \mid v \to m \in R\}$.  In an
usual context free grammar, the sets $M_v$ are finite.  The grammar
$G$ is said to be a {\em generalized context-free grammar} if the
languages $M_v$ are all context-free.

Derivations are defined as usual. More precisely, given $v, w \in (A
\cup V)^*$, we denote by $v \to w$ the fact that $v$ {\em
  directly derives} $w$ and by $v \stackrel{*}{\to} w$ the
fact that $v$ {\em derives} $w$.  The \emph{language generated} by
$G$, denoted $\lang{G}$, is the set of words over $A$ derived from
$S$, i.e., $\lang{G} = \{ u \in A^* \mid S \stackrel{*}{\to} u
\}$.
 
It will be convenient, in the sequel, to use the notation
$v\to\sum_{m\in M_v}m$ or $v \to M_v$ as shortcuts for the set of
rules $\{ v \to m \mid m \in M_v\}$.

Thus, the only difference between usual and generalized context-free
grammars is that for the latter the set of productions may be
infinite, and in this case it is itself context-free.

\begin{theo} \label{subst2} {\upshape{\cite{Kral70}}}
  The language generated by a generalized context-free grammar is
  context-free.
\end{theo}
A sketch of the proof of this theorem is given  in the appendix.

\subsection{Pure alphabetic splicing systems}

A splicing rule $r= \rul{\alpha}{\beta}{\gamma}{\delta}$ is
\emph{pure} if both $\alpha$ and $\beta$ are nonempty. If the rule is
alphabetic, this means that $\alpha$ and $\beta$ are letters. A
splicing system is pure if all its rules are pure.

\begin{theo} \label{Insertion} The language generated by an alphabetic
  context-free pure splicing system is context-free.
\end{theo}

\begin{preuve}
Let ${\cal S} =(A,I, R)$ an alphabetic context-free pure
system. We suppose that the set $R$ is complete.

We construct a generalized context-free  grammar $G$ with axiom $S$,
terminal alphabet $A$  and  with non-terminals $S$ and 
$\ins{a}{b}, \wrdb{a}{b}$ for $a,b\in A$, and $\singb{a} $ for $a\in A$.

The variable $\wrdb{a}{b}$ is used  to derive words with at 
least two letters that begin with a letter $a$
 and end with a letter $b$. The variable $\singb{a}$ is used to 
derive the word  $a$ if it is in the set $I$.

A symbol $\ins{a}{b}$ is always preceded by a letter $a$ or by a
letter $\singb{a}$ or by a letter $\wrdb{c}{a}$, and is always followed
by a letter $b$ by or a letter $\singb{b}$ or by a letter
$\wrdb{b}{d}$.  Roughly speaking, the symbol $\ins{a}{b}$ denotes
words for which eventually there is a letter $a$ preceding it and and
a letter $b$ following it.

We define an operation 
\begin{displaymath}
\Ins:A^+\to (A\cup\bigcup_{a,b\in
  A}\ins{a}{b})^+
\end{displaymath}
by $\Ins(x)=x$ for $x\in A$, and on words $a_1a_2a_3\cdots
a_{n-1}a_n$ where $a_1,\dots,a_n\in A$ and $n\ge2$, by setting
\begin{displaymath}
  \Ins(a_1a_2a_3\cdots a_{n-1}a_n) 
  = a_1\; \ins{a_1}{a_2}\; a_2\; \ins{a_2}{a_3}\; a_3  
  \ldots a_{n-1}\;\ins{a_{n-1}}{a_n} \;a_n\,.
\end{displaymath}
The derivation rules of $G$ are divided into the three following
groups. (Here
$a$ and $b$ are letters in $A$.)

The first group  contains derivation rules that separate  words
according to their initial and final letters, and single out
one-letter words.
\begin{align*}
  &S \to\wrdb{a}{b}\,,\\
  &S\to \singb{a}\,,\\
  &\wrdb{a}{b}  \to \Ins( I \cap aA^*b)\,,\\
  &\singb{a} \to I\cap a\,.
\end{align*}
We use here the convention that a derivation rule of the last type is not
added if $I\cap a$ is empty. Similarly, the third sets in these
derivation rules may be empty. Observe that these sets may also be
context-free.

\noindent The second group  reflects the application of the rules in $R$. It is
composed of
\begin{align*}
&\ins{a}{b}\to \ins{a}{c}\;\wrdb{c}{d}\;\ins{d}{b}\,,\quad\text{for}\ 
 \rul{a}{b}{c}{d} \in R\,,\\
&\ins{a}{b}\to \ins{a}{c}\; \singb{c}\; \ins{c}{b}\,,\quad\text{for}\ 
\rul{a}{b}{c}{\e}\in R\ \text{or}\ 
\rul{a}{b}{\e}{c} \in R\,. 
\end{align*}

\noindent The third group  of derivation rules is used to  replace the variables
$\ins{a}{b}$ by the empty word. 
\begin{align*}
\ins{a}{b} \to \e\,.
\end{align*}

\noindent By Theorem \ref{subst2}, the language generated by $G$ is
context-free. 

We claim that $\lang{G} = \langf{{\cal S}}$.  
Consider a derivation
\begin{displaymath}
S\stackrel{*}{\rightarrow}w\,,\quad\text{with $w\in A^+$}
\end{displaymath}
in the grammar $G$. Suppose now that, in this derivation, we remove all
derivation steps involving a derivation rule of the third group. Then
the derivation is still valid, and the result is a derivation
\begin{displaymath}
  S\stackrel{*}{\rightarrow}\Ins(w)\,.
\end{displaymath}
Conversely, given  a derivation $S\stackrel{*}{\rightarrow}\Ins(w)$, one gets a derivation
$S\stackrel{*}{\rightarrow}w$ by simply applying the necessary
derivation rules of the third group.

We denote by $\langprime{G}$ the language obtained without applying the
productions of the third type, and by $\sublangprime{G}{\wrdb{a}{b}}$
and by $\sublangprime{G}{\singb{a}}$ the languages obtained when
starting with the variable $\wrdb{a}{b}$ (resp. with $\singb{a}$),
and we prove that $\langprime{G}=\Ins(\langf{{\cal S}})$.

First, we prove the inclusion $\langprime{G} \subseteq
\Ins(\langf{{\cal S}})$.  For this, we prove by induction on the
length of the derivations in $G'$ that for all letters $a, b \in A$,
we have $ \sublangprime{G}{\wrdb{a}{b}}\subseteq \Ins(\langf{{\cal
    S}}\cap aA^*b)$ and that $ \sublangprime{G}{\singb{a}}\subseteq
\Ins(\langf{{\cal S}}\cap a)$.

A derivation $X\stackrel{*}{\rightarrow} v$ is called \emph{terminal}
if $v$ does not contains any occurrence of variables other than
$\ins{a}{b}$, for $a,b\in A$.
It is easy to check that the length of terminal derivations are always odd.
 
The only terminal derivations of length one are
\begin{align*}
&\wrdb{a}{b}  \to \Ins( I \cap aA^*b)\,,\\
&\singb{a} \to I \cap a\,,
\end{align*}
and the inclusion is clear.

Assume that the hypotheses of induction hold for derivations of length
less than $k$ and let $u$ be a word obtained by a derivation of
length~$k$.  Since the length of the derivation is greater than $1$,
the derivation starts with a derivation step $S\to \wrdb{a}{b}$ for
some $a,b\in A$. The last two derivation steps have one of the
following form
\begin{align*}
&\ins{c}{d}\to \ins{c}{e}\: \wrdb{e}{f}\: \ins{f}{d}
  \to \ins{c}{e}\:x \: \ins{f}{d}\,,\quad\text{with $x \in \Ins( I \cap
  eA^*f )$}\,,\\
&\ins{c}{d}\to \ins{c}{e}\: \singb{e}\: \ins{e}{c}
  \to \ins{c}{e}\: e \: \ins{e}{c}\,,\quad\text{with $e \in I$}\,,
\end{align*}
for suitable letters
$c, d, e, f$.

In the first case, there are words $v,w$ such that
\begin{displaymath}
\wrdb{a}{b} \stackrel{*}{\rightarrow} v \ins{c}{d} w
\to  v \ins{c}{e} \wrdb{e}{f} \ins{f}{d} w
\to  v \ins{c}{e} x \ins{f}{d} w = u\,.
\end{displaymath}
By induction, $v \:\ins{c}{d}\: w \in \Ins(\langf{{\cal S}}\cap
aA^*b)$.  Since the derivation rule $\ins{c}{d}\to \ins{c}{e}\:
\wrdb{e}{f}\: \ins{f}{d}$ is in $G$, there is a splicing rule $\rul c
d e f$ in $R$.  Consequently, the word $u$ is in $\Ins( \langf{{\cal
    S}} \cap aA^*b)$.  The second case is similar.  This proves the
inclusion $\langprime{G} \subseteq \Ins(\langf{{\cal S}})$.

Now, we prove the inclusion $ \Ins(\langf{\cal S}) \subseteq
\langprime{G}$.  For this, we prove that for all letters $a, b \in A$, we
have $\Ins( \langf{{\cal S}} \cap aA^*b) \subseteq
\sublangprime{G}{\wrdb{a}{b}}$ and $\Ins( \langf{{\cal S}} \cap a)
\subseteq \sublangprime{G}{\singb{a}}$
 
We observe that for a letter $a$, one has $\langf{{\cal S}} \cap
a=I\cap a= \Ins( \langf{{\cal S}} \cap a)$.  The letter $a$ is thus
obtained by the derivation $\singb{a} \to I \cap a$.  Thus we have
$\Ins( \langf{{\cal S}} \cap a) \subseteq \sublangprime{G}{\singb{a}}$
for all letters $a \in A$.

Let us prove the inclusions $\Ins( \langf{{\cal S}} \cap aA^*b) \subseteq
\sublang{G'}{\wrdb{a}{b}}$ by induction on the number of splicing
rules used for the production of a word in $\langf{{\cal S}} \cap
aA^*b$. 

Let $u \in \langf{{\cal S}} \cap aA^*b$. If no splicing rule is used,
then $u \in I \cap aA^*b$. The word $\Ins(u)$ is obtained by the
application of the corresponding derivation rule $\wrdb{a}{b}
\to\Ins(u)$ which is in the set $\wrdb{a}{b} \to \Ins( I \cap
aA^*b)$. Thus $u \in \sublangprime{G}{\wrdb{a}{b}}$.

Assume that the inductive hypothesis holds for the words obtained by
less than $k$
splicing operations, and that
$u$ is  obtained by application of $k\ge1$ splicing operations. We
consider the last  insertion that leads to $u$:
there exist three nonempty words 
$v$, $w$ and $x$ and a pure rule $r\in R$, such that 
$\produc{r}{v\cdot w}{x}{u = v\cdot x\cdot w }$, and moreover $vw$ and $x$ are 
words of  $\langf{{\cal S}}$ obtained by less than $k$
splicing operations

Two cases may occur, for suitable letters $e$ and $f$:
\begin{align*}
&x \in \langf{{\cal S}} \cap eA^*f\,,\\
&x \in \langf{{\cal S}} \cap e\,.
\end{align*}
Consider the first case. Let $c$ be the last letter of $v$, and let
$d$ be the first letter of $w$. Then $r=\rul c d e f$.
By induction hypothesis, we have 
$\wrdb{a}{b} \stackrel{*}{\to} \Ins(v) \:\ins{c}{d} \:\Ins(w)\:(= \Ins(vw))$ 
and $\wrdb{e}{f} \stackrel{*}{\to} \Ins(x)$.
Moreover, the rule $r$ shows that the derivation rule
 $\ins{c}{d}\to \ins{c}{e}\: \wrdb{e}{f}\: \ins{f}{d}$ is in the
 grammar $G$. 
Thus combining these three derivations, we obtain
\begin{align*}
  \wrdb{a}{b} 
  &\stackrel{*}{\to} \Ins(v) \:\ins{c}{d} \:\Ins(w) 
  \to  \Ins(v) \:\ins{c}{e}\: \wrdb{e}{f}\: \ins{f}{d}\:\Ins(w)\\ 
&\stackrel{*}{\to}  \Ins(v) \:\ins{c}{e}\:\Ins(x)\: \ins{f}{d}\:\Ins(w)\,.
\end{align*}
Thus $\Ins(u) \in \sublangprime{G}{\wrdb{a}{b}}$.
The second case is similar.

This shows the inclusion $\Ins(\langf{{\cal S}}) \subseteq
\langprime{G}$. Consequently $\Ins(\langf{{\cal S}}) = \langprime{G}$, and
quite obviously, we can deduce $\langf{{\cal S}} = \lang{G}$.
\end{preuve}

\begin{example}\label{ex-pure}
  Consider the pure splicing system 
  \begin{displaymath} 
    {\cal S} = (A, I, R)
  \end{displaymath}
  with $A = \{ a, b, c\}$, $I = c^*ab\cup c$, and with $R$ composed of
  the rules
  \begin{displaymath}
    r = \rul{c}{b}{\e}{a}\,,\ r' = \rul{c}{c}{\e}{b}\,,\ 
    r''= \rul{a}{b}{a}{b}\,.
  \end{displaymath}
  This splicing system generates the language $\langf{{\cal S}} =
  c(c\cup L)^+L\cup\{c\}$, with $L = \{ a^nb^n \mid n \geq 1\}$.

  For the construction of the grammar for $\langf{{\cal S}}$, we add
  the completions of the rules $r$ and $r'$. We also discard tacitly
  useless variables.  Now, we observe that $I\cap aA^*b=ab$, $I\cap
  cA^*b=c^+ab$, $I\cap c=c$, and that the other intersections are
  empty. Thus, the first group of derivation rules of the grammar is
  the following.

\begin{align*} 
  S & \to \wrdb{a}{b}\mid  \wrdb{c}{b} \mid \singb{c}\\
  \wrdb{a}{b} & \to  a \:\ins{a}{b}\:b\\
  \wrdb{c}{b} & \to  (c \:\ins{c}{c})^* c \:\ins{c}{a}\:a \:\ins{a}{b}\:b\\
  \singb{c}  & \to   c
\end{align*}
We observe by
inspection, that there is no derivation rule starting with
$\wrdb{b}{x}$, $\wrdb{x}{a}$ or $\wrdb{x}{c}$, for $x\in A$, and
similarly for $\singb{a},\singb{b}$.  This leaves only the following second
group of rules.
\begin{align*} 
   \ins{a}{b}& \to \ins{a}{a}\: \wrdb{a}{b}\: \ins{b}{b}\\
  \ins{c}{a}& \to \ins{c}{a}\: \wrdb{a}{b}\: \ins{b}{a}\\
  \ins{c}{a}& \to \ins{c}{c}\: \wrdb{c}{b}\: \ins{b}{a}\\
  \ins{c}{c}& \to \ins{c}{a}\: \wrdb{a}{b}\: \ins{b}{c}\\
  \ins{c}{c}& \to \ins{c}{c}\: \wrdb{c}{b}\: \ins{b}{c}
\end{align*}
When looking for the final grammar, we may observe that the variables
$\ins{b}{x}$ for $x\in A$, and $\ins{a}{a}$ only produce the empty
word. Also they can be replaced by $\e$ everywhere in the grammar. It
follows that $\ins{a}{b}$ can be replaced by $\wrdb{a}{b}$. Also, it
is easily seen that $\ins{c}{a}$ and $\ins{c}{c}$ generate the same
language. This leads to the following grammar, where we write, for
easier reading, $X$ for $\wrdb{a}{b}$ and $Y$ for $\wrdb{c}{b}$, and
$T$ for $\ins{c}{a}$.
\begin{align*} 
  S & \to X\mid  Y \mid c\\
  X & \to aXb \mid ab\\
  Y & \to (cT)^+X\\
  T & \to TX\mid TY\mid \e
\end{align*}
It is easily checked that this generalized context-free grammar indeed
generates the language $\langf{{\cal S}} = c(c\cup L)^+L\cup\{c\}$,
with $L = \{ a^nb^n \mid n \geq 1\}$.
\end{example}


\section{Concatenation systems\label{sec-echange}}



We introduce a classification of the productions generated in a
splicing system by defining two kinds of productions called proper
insertions and concatenations.

Let $r= \rul{\alpha}{\beta}{\gamma}{\delta}$ be a splicing rule.  The
production $\produc{r}{x\alpha \cdot \beta y}{\gamma z
  \delta}{x\alpha\cdot \gamma z \delta\cdot \beta y}$ is a {\em proper
  insertion} if $x\alpha \neq \e$ and $\beta y \neq \e$, it is a {\em
  concatenation} otherwise. If $r$ is a pure rule, then its
productions are always proper insertions.

Of course, the rule $r$ can produce a concatenation only if $\beta =
\e$ or $\alpha = \e$.  However, such rules can be used for both kinds
of productions.  Consider for example the rule $r
=\rul{a}{\e}{c}{d}$. Then the production $\produc{r}{aa\cdot }{cad}{aa
  \cdot cad}$ is a concatenation, while the production
$\produc{r}{a\cdot a}{cad}{a\cdot cad \cdot a}$ is a proper insertion.
We consider now rules which are not pure, and we restrict their usage to
concatenations. This leads to the notion of concatenation systems.  We
then show that alphabetic context-free concatenation systems only
generate context-free languages.

\subsection{Concatenation systems}

A {\em concatenation system}   is a triplet ${\cal T}=(A,I, R)$, where
$A$ is an alphabet, $I$ is a set of words over $A$, called the {\em
  initial set} and $R$ is a finite set of {\em concatenation rules}.
A concatenation rule $r$ is a
quadruplet of words
over $A$. It is denoted $r= \rulc{\alpha}{\beta}{\gamma}{\delta}$, to
emphasize the special usage which is made of such a rule.

A concatenation rule $r = \rulc{\alpha}{\beta}{\gamma}{\delta}$ can be
applied to words $u$ and $v$ provided $u\in \alpha A^*\beta$ and $v
\in \gamma A^* \delta$. Applying $r$ to the pair $(u,v)$ gives the
word $w=uv$. This is denoted by $\produconc{r}{u}{v}{w}$ and is called
a {\em concatenation production}.

The \emph{language generated} by the system ${\cal T}$, denoted by
$\langconc{\cal T}$, is the smallest language containing $I$ and
closed under the application of the rules of $R$.

Again, the system ${\cal T}$ is alphabetic if every rule in $R$ have
handles of length at most one.  It is context-free if the initial set
$I$ is context-free.  The notion of complete set is similar to the one
for splicing rules.


\subsection{Alphabetic concatenation}\label{section-concatenation}


This section is devoted to the proof of  the following theorem.

\begin{theo} \label{Concatenation} The language generated by an
  alphabetic context-free concatenation system is context-free.
\end{theo}

\begin{preuve} 
  Let ${\cal T} =(A,I, R)$ an alphabetic context-free concatenation
  system. We suppose that the set $R$ is complete. Set $K=\langconc{{\cal T}}$.

  We construct a grammar $G= (T,V,S,R)$ and a substitution
  $\sigma:T^*\to A^*$ for which we prove that $K = \sigma(\lang{G})$.
  The grammar is quite similar to that built for
  Theorem~\ref{Insertion}.  The grammar $G$ has the set of terminal
  symbols $T = \{\wrdi{a}{b}\mid a,b\in A\}\cup\{\singi{a}\mid a\in
  A\}$, and the set of non-terminal symbols $V = \{S\}
  \cup\{\wrd{a}{b}\mid a, b\in A\}\cup\{\sing{a}\mid a\in A\}$. The
  axiom is $S$.

  As in the proof of Theorem~\ref{Insertion}, the purpose of the
  variables is the following.  The symbol $\wrd{a}{b}$ is used to
  derive words of length at least~$2$ that start with the letter $a$
  and end with the letter $b$, that is the set $K \cap aA^*b$.
  Similarly, the symbol $\sing{a}$ will be used to derive the word $a$
  if it is in $K$.  The terminal symbols $\wrdi{a}{b}$ (resp.
  $\singi{a}$) are mapped to the sets $I \cap aA^*b$ (resp.  $I \cap
  a$) by the context-free substitution $\sigma$ defined by:
\begin{displaymath}
 \sigma( \wrdi{a}{b}) = I \cap aA^*b\,;\quad  \sigma(\singi{a}) = I \cap a\,.
\end{displaymath}
This substitution is context-free because the set $I$ is context-free. 

The derivation rules of the grammar $G$ are divided in two groups.
In the following, $a$ and $b$ are any letters in $A$.

The first group  
contains derivation rules that separate  words
according to their initial and final letters, and single out
one-letter words:
\begin{align*}
&S \to \wrd{a}{b}\,,\\
&S\to \sing{a}\,,\\
& \wrd{a}{b} \to \wrdi{a}{b}\,,\\
&\sing{a}\to \singi{a}\,.
   \end{align*}
The second group of rules  deals with concatenations:
\begin{enumerate}[$\qquad\quad\ $]
\item $\wrd{a}{b} \to\wrd{a}{c}\ \wrd{d}{b}\,,\quad$ for
  $\rulc{a}{c}{d}{b} \in R$,,
  \item$\wrd{a}{b} \to\sing{a}\   \wrd{c}{b}\,,
  \quad$for  $\rulc{\e}{a}{c}{b}  \in R$
    or  $\rulc{a}{\e}{c}{b} \in R$\,,
  \item$\wrd{a}{b} \to\wrd{a}{c}\ \sing{b}\,, 
  \quad$for
    $\rulc{a}{c}{\e}{b}  \in R$
    or  $\rulc{a}{c}{b}{\e} \in R$\,,
  \item$\wrd{a}{b} \to\sing{a}\   \sing{b}\,, 
  \quad$for   
    $\rulc{\e}{a}{\e}{b}\in R$
    or  $\rulc{a}{\e}{\e}{b} \in R$
  or $\rulc{a}{\e}{b}{\e} \in~R$
    or  $\rulc{\e}{a}{b}{\e} \in R$\,.
\end{enumerate} 
By construction, the language $\lang{G}$ generated by $G$ is
context-free, and by Theorem \ref{subst1}, the language
$\sigma(\lang{G})$ is also context-free.

We claim that $ \sigma(\lang{G}) =K$.  We first prove the inclusion $
\sigma(\lang{G}) \subseteq K$.  For this, we show, by induction on the
length of the derivation in $G$, that for all letters $a, b \in A$, we
have $ \sigma(\sublang{G}{\wrd{a}{b}})\subseteq K
\cap aA^*b$ and that $ \sigma(\sublang{G}{\sing{a}})\subseteq
K\cap a$.

The only terminal derivations of length~$1$ are
\begin{align*}
&\wrd{a}{b} \to \wrdi{a}{b}\text{\ and one has $\sigma(\wrdi{a}{b}) =
  I \cap aA^*b \subseteq K\cap aA^*b$}\,,\\
&\sing{a}\to \singi{a}\text{\  and one has $\sigma(\singi{a})=I\cap
  a\subseteq K\cap a$}\,.
\end{align*}
Thus the inclusion holds in this case.

Assume that the hypotheses of induction are true for derivations of
length less that $k$ and let $u$ a word obtained by a derivation of
length $k$.  Since $k\ge2$, the first derivation rule is one of the
second group, and the
derivation has one of the forms
\begin{align*}
  \wrd{a}{b}&\to \wrd{a}{c}\:\wrd{d}{b}\stackrel{*}{\to} u\\
  \wrd{a}{b}&\to \sing{a}\:\wrd{c}{b}\stackrel{*}{\to} u\\
  \wrd{a}{b}&\to \wrd{a}{c}\:\sing{b}\stackrel{*}{\to} u\\
  \wrd{a}{b}&\to \sing{a}\,\sing{b}\stackrel{*}{\to} u
\end{align*}
for some $a,b,c,d\in A$.  In the first case, we have $u = u_1 u_2$,
with $\wrd{a}{c} \stackrel{*}{\to} u_1$ and $\wrd{d}{b}
\stackrel{*}{\to} u_2$, both derivations having length strictly less
than $k$. By the inductive hypotheses, $\sigma(u_1) \in K \cap aA^*c$
and $\sigma(u_2) \in K \cap dA^*b$.  Moreover, since $\wrd{a}{b}
\to\wrd{a}{c}\ \wrd{d}{b}$ is a derivation rule in $G$, one has
$\rulc{a}{c}{d}{b} \in R$. This ensures that $(K \cap aA^*c)(K \cap
dA^*b)\subseteq K \cap aA^*b$.  Consequently, $\sigma(u)= \sigma(u_1)
\sigma(u_2)$ is in $ K\cap aA^* b $.  The other cases are similar.
This proves the inclusion $ \sigma(\lang{G}) \subseteq K$.

We now prove the converse inclusion $ K \subseteq \sigma(\lang{G})$.
For this, we prove that for all letters $a, b \in A$, we have $K \cap
aA^*b \subseteq \sigma(\sublang{G}{\wrd{a}{b}})$ and that $K \cap a
\subseteq \sigma(\sublang{G}{\sing{a}})$.

It is easy to see, that if $ a \in K$ then $a\in \singi{a}$, $\sigma(
\singi{a}) = {a}$. and $\sing{a} \to \singi{a}$.  Thus $K \cap a
\subseteq \sigma(\sublang{G}{\sing{a}})$, for all letter $a$ in $A$.

The inclusions $K \cap aA^*b \subseteq
\sigma(\sublang{G}{\wrd{a}{b}})$  are proved
by induction on the number of the concatenation operations used. 
Let $u\in K \cap aA^*b$.

If $u$ is obtained without any concatenation, then $u \in I \cap
aA^*b=\sigma(\wrdi{a}{b})$,
and since $ \wrd{a}{b} \to \wrdi{a}{b}$ is a derivation rule in $G$,
we have $u \in  \sigma(\sublang{G}{\wrd{a}{b}})$.

Assume that the inductive hypothesis holds for  words obtained by
less than $k$ concatenations, and that $u$ is  obtained by  $k$
concatenations.
Then there exist two words 
$u_1$ and $u_2$ such that $u = u_1 u_2$ and such that $u_1$ and $u_2$
are obtained by less than $k$ concatenations.
There are four cases to consider, according to the concatenation rule
uses to produce $u$ from $u_1$ and $u_2$. The cases are the following.
\begin{align*}
&u_1 \in K \cap aA^*c\,,\quad  u_2 \in K \cap dA^*b\,,\\
&u_1 \in K \cap aA^*c\,,\quad u_2 \in K \cap b\,,\\
&u_1 \in K \cap a\,,\quad u_2 \in K \cap dA^*b\,,\\
&u_1 \in K \cap a\,,\quad u_2 \in K \cap b\,.
\end{align*}
Consider the first case (the other are similar). Since $u\in K$, there
is a concatenation rule $\rulc{a}{c}{d}{b}$ in $R$. Consequently,
there exists in $G$  a derivation rule $\wrd{a}{b}  \to \wrd{a}{c} \: \wrd d{b}$.
By induction hypothesis, there is a derivation 
$\wrd{a}{c} \stackrel{*}{\to}
v_1$ with  $u_1 = \sigma(v_1)$, and a derivation 
$\wrd{d}{b} \stackrel{*}{\to}
v_2$ with  $u_2 = \sigma(v_2)$. It follows that
\begin{displaymath}
  \wrd{a}{b} \to \wrd{a}{c}\;\wrd{d}{b} \stackrel{*}{\to}v_1v_2\,,
\end{displaymath}
and since $\sigma( v_1 v_2) = \sigma(v_1)\sigma(v_2) = u$, one has $u
\in \sigma(\sublang{G}{\wrd{a}{b}})$.  This proves the inclusion $ K
\subseteq \sigma(\lang{G})$, and thus the claim. Since
$\sigma(\lang{G})$ is context-free, the language $K$ is also
context-free. This completes the proof.
\end{preuve}

\begin{rema}
  Contrary to Theorem~\ref{Insertion} which is false for systems which
  are not alphabetic, Theorem~\ref{Concatenation} holds for
  concatenation systems without the requirement that they are
  alphabetic. The proof is quite analogous to the alphabetic case.
\end{rema}

\begin{example}\label{ex-concat}
  Consider the concatenation system ${\cal T} = (A, I, R)$ over the
  alphabet $A = \{ a, b, c\}$, with $I = \{ab, c\}$, and with $R$ composed of
  the concatenation rules 
   \begin{align*}
       &\rulc{\e}{c}{\e}{b}\,, \\
       &\rulc{\e}{c}{x}{b}\quad\text{for $x\in A$}\,.
   \end{align*}
The completion of the system gives the concatenation rules
\begin{align*}
        &\rulc{\e}{c}{\e}{b}\,, \\
        &\rulc{\e}{c}{x}{b}\quad\text{for $x\in A$}\\
        &\rulc{y}{c}{\e}{b}\quad\text{for $y\in A$}\\
        &\rulc{y}{c}{x}{b}\quad\text{for $x,y\in A$}\,.
\end{align*}
According to
the construction of the previous proof, these concatenation rules give
the derivation rules
\begin{align} 
 \wrd cb&\to\sing c\ \sing b\label{(1)}\\
 \wrd cb&\to\sing c\ \wrd xb\quad\text{for $x\in A$}\label{(2)}\\
 \wrd yb&\to\wrd yc\ \sing b\quad\text{for $y\in A$}\label{(3)}\\
 \wrd yb&\to\wrd yc\ \wrd xb\quad\text{for $x,y\in A$}\label{(4)}
\end{align}
The first group of
derivation rules is composed only of
\begin{align*} 
  S           &\to \wrd{x}{y}\quad\text{for $x,y\in A$}\\
  S           &\to \sing{c}\\
  \wrd{a}{b}  &\to  \wrdi{a}{b}\\
  \sing{c}    &\to  \singi{c}
\end{align*}
because of the set $I$ of initial words.
Since there is no derivation rule starting with $\sing b$, the
derivation rules~\eqref{(1)} and~\eqref{(3)} are useless and can be
removed. Similarly, there is no derivation rule starting with $\wrd
yc$, so the dervation rules~\eqref{(4)} can be removed. For the same
reason, the variable  $\wrd bb$  can be removed. Finally, we get
the grammar 
\begin{align*} 
  S           &\to \wrd{a}{b} \mid \wrd{c}{b} \mid \sing{c}\\
  \wrd{a}{b}  &\to  \wrdi{a}{b}\\
  \sing{c}    &\to  \singi{c}\\
  \wrd{c}{b}  & \to  \sing{c}\   \wrd{a}{b}\mid  \sing{c}\   \wrd{c}{b}
\end{align*}
and the substitution
\begin{align*} 
\sigma(\wrdi{a}{b}) &=ab\\
 \sigma(\singi{c}) &= c
\end{align*}
The language obtained is $c^*ab+c$.
\end{example}



\begin{rema}
  The language $\langconc{\cal T}$ generated by a concatenation system
  ${\cal T} = (A, I, R)$ may not be regular, even if $I$ is
  finite. Consider indeed the system given by $I = \{ab,a,b,c,d\}$ and
  \begin{displaymath}
    R = \{ \rulc{\e}{c}{a}{b}, \rulc{c}{b}{d}{\e}, \rulc{\e}{a}{c}{d},
    \rulc{a}{d}{b}{\e} \}\,.
  \end{displaymath}
  The language obtained is $\langconc{{\cal T}} = L \cup cL \cup cLd
  \cup acLd$ where $L$ denotes the $L=\{ (ac)^nab (db)^n \mid n\geq 0
  \}$, and this language is not regular.
\end{rema}

\subsection{Heterogeneous systems}

A splicing system is a {\em heterogeneous system}
if all its rules are either pure rules or concatenation rules.

The aim of heterogeneous systems is to separate the splicing rules
according the their usage. A pure rule is used for a proper insertion,
that is for producing a word $w=xvy$ from words $u=xy$ and $v$, with
$x,y\ne\e$. On the contrary,  a concatenation rule produces the word
$w=uv$ or $w=vu$, that is handles precisely the case where $x=\e$ or
$y=\e$.

The following proposition shows that for any  flat alphabetic splicing system, there is an alphabetic heterogeneous system
  with same initial set $I$ which generates the same language.

\begin{prop} \label{conversion} 
  Let ${\cal S} = (A, I, R)$ be a complete alphabetic splicing system,
  and let ${\cal S'} = (A, I, R'\cup R'')$ be the heterogeneous system
  with same initial set $I$, where $R'$ is the set of pure rules of
  $R$, and
  \begin{displaymath}
R''=\{\rulc{\e}{\alpha}{\gamma}{\delta}\mid
  \rul{\alpha}{\e}{\gamma}{\delta} \in
  R\}\cup\{\rulc{\gamma}{\delta}{\beta}{\e}\mid
  \rul{\e}{\beta}{\gamma}{\delta} \in R\}\,.
\end{displaymath}
Then $S$ and $S'$ generate
  the same language.
\end{prop} 

\begin{preuve}
%
%
%
%
The verification
is left to the reader.
\end{preuve}

\begin{example}\label{ex-homogeneous}
Let ${\cal S}$ be the flat splicing system $(A, I, R)$ with $A = \{ a, b, c\}$, 
$I = \{ab, c\}$, and $R = \{ \rul{a}{b}{a}{b},\rul{c}{\e}{\e}{b}\}$.

We complete $R$. 
The complete set of rules for $R$ is
\begin{displaymath}
     \begin{array}{rl}
     \rul{a}{b}{a}{b} & \\
     \rul{c}{\e}{\e}{b}&\\
     \rul{c}{y}{x}{b} & \text{ for } x,y \in \{ a, b, c\}\\
      \rul{c}{\e}{x}{b} & \text{ for } x \in \{ a, b, c\}\\
      \rul{c}{x}{\e}{b} & \text{ for } x \in \{ a, b, c\}
      \end{array}
  \end{displaymath}
The heterogeneous system ${\cal S'}$ corresponding to ${\cal S}$
is the system ${\cal S'} = (A, I, R')$
with $R'$ is composed of the pure rules
\begin{displaymath}
     \begin{array}{rl}
     \rul{a}{b}{a}{b} & \\
     \rul{c}{y}{x}{b}  & \text{ for } x,y \in \{ a, b, c\}\\
      \rul{c}{x}{\e}{b} & \text{ for } x \in \{ a, b, c\}
      \end{array}
  \end{displaymath}
and with the concatenation rules
 \begin{displaymath}
     \begin{array}{ll}
      \rulc{\e}{c}{x}{b} & \text{ for } x \in \{ a, b, c\} \\
      \rulc{\e}{c}{\e}{b} &
      \end{array}
  \end{displaymath}
which, after completion, give the concatenation rules of
Example~\ref{ex-concat}.
\end{example}

\subsection{Weak commutation of concatenations and proper insertions}

Given a heterogeneous system ${\cal S}= (A,I, R)$, a {\em production sequence}
is a sequence $[\pi_1;\pi_2;\dots;\pi_n]$ of productions such that,
setting
\begin{displaymath}
  \pi_k\ =\ (\produc{r_k}{u_k}{v_k}{w_k})\quad\text{for $1\le k\le n$,}
\end{displaymath}
each $u_k$ and $v_k$ is either an element of $I$, or is equal to one
of the words $u_1,v_1,w_1,\ldots,u_{k-1},v_{k-1},w_{k-1}$.
The word $w_n$ is the {\em result} of the production sequence.
The length of the sequence is $n$.
By  convention,  there is a production sequence of length~$0$ for each
$w\in I$, denoted by $[w]$. Its result is $w$. 

\begin{example} \label{seq_ex}
Consider the  pure system over $A  = \{a, b\}$ with initial set 
$I= \{ab\}$ and the unique splicing rule $r =\rul{a}{b}{a}{b}$. 
In this system, the only  splicing sequence  of length 0 is $[ab]$.
Both production sequences (we omit the reference to $r$)
\begin{displaymath}
 [\produc{}{ab}{ab}{a^2b^2};\  \produc{}{a^2b^2}{a^2b^2}{a^4b^4}]
\end{displaymath}
and 
\begin{displaymath}
[\produc{}{ab}{ab}{a^2b^2};\  \produc{}{ab}{a^2b^2}{a^3b^3}; \
\produc{}{a^3b^3}{ab}{a^4b^4} ] 
\end{displaymath}
have the same result $a^4b^4$.
\end{example}

Clearly, the language $\langf{{\cal S}}$ generated by a heterogeneous
system $\cal S$ is the set of the results of all its production sequences.

We show that, in an alphabetic splicing system, one always can choose
a particular type of production sequence for the computation of a word,
namely a sequence where the concatenations are performed before proper
insertions. This is stated in the following lemma.

\begin{lem} \label{exchange} Let ${\cal S}=(A,I,R)$ be an alphabetic
  heterogeneous 
  splicing system.  Given a sequence of proper insertions and
  concatenation productions with result $u$, there exists another
  sequence with same result $u$, using the same rules of proper
  insertions and concatenations, and such
  that all concatenation productions occur before any proper insertion
  production.
\end{lem}


\begin{preuve}
Let $r_1 = \rul{\alpha}{\beta}{\gamma}{\delta}$ be a pure
rule and let $r_2 = \rulc{\zeta}{\eta}{\mu}{\nu}$ be a concatenation
rule, and assume that there is production sequence
$\sigma=[\pi_1;\pi_2]$, with
\begin{displaymath}
  \pi_1 =(\produc{r_1}{u}{v}{w})\,,\quad\pi_2 =(\produconc{r_2}{p}{s}{t})\,,
\end{displaymath}
where $u,v,w,p,s,t$ are words and $t=ps$.
We assume that $u,v,p,s$ are all non-empty.
If neither $p$ nor $s$ is equal to $w$ we replace the sequence 
$[\pi_1;\;\pi_2]$ by
by $[\pi_2;\; \pi_1]$, and we get the result.

Assume now that $p=w$ or $s = w$.  Since $\pi_1$ is
a proper insertion, there exists a factorization $u = u_1\cdot u_2$,
with $u_1$ and $u_2$ non-empty words, such that $w = u_1 \cdot v \cdot
u_2$, and the production $\pi_1$ can be rewritten as $\pi_1
=(\produc{r_1}{ u_1\cdot u_2}{v}{u_1 \cdot v \cdot u_2})$.

There are two cases to be considered. 
\begin{enumerate}
\item $p = w, s \neq w$ (or the symmetric case  $p \neq w, s = w$);
\item $p = s = w$.
\end{enumerate}

\noindent Case 1: In this case, we have
\begin{displaymath}
\pi_1 =(\produconc{r_1}{ u_1\cdot u_2}{v}{u_1 \cdot v \cdot u_2})\,, \quad
\pi_2 =(\produins{r_2}{u_1v u_2}{s}{ u_1 v u_2\cdot s })\,,
\end{displaymath}
and, in view of the production $\pi_2$, one has $u_1 \in \zeta A^*$
and $u_2 \in A^*\eta$ (here we use the fact that $u_1$ and $u_2$ are
non-empty, and that $\zeta$ and $\eta$ have at most one letter).
  
We replace the sequence $[\pi_1;\pi_2]$ by 
$[\pi_3;\;\pi_4]$, where
\begin{displaymath}
\pi_3 =(\produconc{r_2}{u_1 u_2}{s}{ u_1 u_2\cdot s})\,,\quad
\pi_4 =(\produc{r_1}{u_1\cdot u_2s}{v}{t = u_1 \cdot v \cdot u_2s})\,.
\end{displaymath}
The  concatenation  $\pi_3$  is valid because and $s \in \mu A ^* \nu$.

\noindent 
Case 2: In this case, 
\begin{align*}
  &\pi_1 =(\produc{r_1}{ u_1\cdot u_2}{v}{u_1 \cdot v \cdot u_2})\,,\\
  &\pi_2 =(\produconc{r_2}{ u_1  v  u_2}{u_1  v  u_2}{u_1 v u_2 \cdot u_1  v  u_2})\,,
\end{align*}
and the sequence $[\pi_1;\;\pi_2]$ 
is replaced by $[\pi_3;\;\pi_4;\;\pi_5;\;\pi_1]$, where
\begin{align*}
    &\pi_3 =(\produconc{r_2}{ u_1   u_2}{u_1   u_2}{ u_1  u_2 \cdot u_1   u_2})\,, \\
   & \pi_4 =(\produc{r_1}{ u_1\cdot u_2u_1u_2}{v}{u_1 \cdot v \cdot u_2u_1u_2})\,, \\
    &\pi_5 =(\produc{r_1}{ u_1  v  u_2 u_1\cdot u_2}{v}{ u_1  v  u_2u_1 \cdot v \cdot u_2})\,.
\end{align*}
This proves the lemma.
\end{preuve}

\begin{rema}
  The proposition does not hold anymore if the rules are not
  alphabetic.
  Consider for example the rules $r_1=\ruli{a}{b}{x}{y}$ and
  $r_2=\rulc{z}{t}{ax}{\e}$.  Then the splicing sequence
  \begin{displaymath}
    [\produc{r_1}{a\cdot b}{x y}{a\cdot xy \cdot b} ;\ 
    \produconc{r_2}{zt}{axyb}{zt\cdot axyb}]
  \end{displaymath}
  cannot be replaced by a sequence where proper insertions occur after
  concatenations.
\end{rema}

The following theorem is an immediate corollary of the previous lemma.

\begin{prop} \label{heterogeneous} 
  For any language $L$ generated by an alphabetic heterogeneous system
  ${\cal S}=(A,I,R)$, there exist a set of alphabetic concatenation
  rules $R'$ and a set of pure alphabetic rules $R''$,
  such that
  \begin{displaymath}
    L = \langf{(A,\langconc{(A,I, R')}, R'')}\,.
  \end{displaymath}
\end{prop}

The combination of Theorems~\ref{Concatenation} and~\ref{Insertion}
gives the following theorem, which is our main theorem in the case
of a flat system.

\begin{theo}\label{theoMainPlat}
  Let ${\cal S}=(A,I,R)$ be a flat alphabetic context-free splicing
  system.  Then $\langf{\cal S}$ is context-free.
\end{theo}

\begin{preuve}
  By Theorem~\ref{heterogeneous}, $\langf{\cal
    S}=\langf{(A,\langconc{(A,I, R')}, R'')}$. The language
  $L=\langconc{(A,I, R')}$ is context-free in view of
  Theorem~\ref{Concatenation}. The language $\langf{(A,L, R'')}$ is
  context-free by Theorem~\ref{Insertion}. Thus $\langf{\cal S}$ is
  context-free. 
\end{preuve}

\begin{example}\label{ex-final}
  Consider again the splicing system ${\cal S}=(A, I, R)$ with $A = \{
  a, b, c\}$, $I = \{ab, c\}$, and $R = \{
  \rul{a}{b}{a}{b},\rul{c}{\e}{\e}{b}\}$.
The homogeneous system corresponding to ${\cal S}$ is given in
Example~\ref{ex-homogeneous}. The associated concatenation system
${\cal T}=(A,I,R')$ has the concatenation rules $R'$ composed of
\begin{displaymath}
  \begin{array}{ll}
    \rulc{\e}{c}{x}{b} & \text{ for } x \in \{ a, b, c\} \\
    \rulc{\e}{c}{\e}{b} &
  \end{array}
\end{displaymath}
and we have seen in Example~\ref{ex-concat} that it generates the
language $\langconc T=c^*ab\cup c$. The pure system has the set $R''$
of rules consisting in
\begin{displaymath}
  \begin{array}{rl}
    \rul{a}{b}{a}{b} & \\
    \rul{c}{y}{x}{b}  & \text{ for } x,y \in \{ a, b, c\}\\
    \rul{c}{x}{\e}{b} & \text{ for } x \in \{ a, b, c\}
  \end{array}
\end{displaymath}
As seen in Example~\ref{ex-pure}, it generates the context-free
language $\langf{{\cal S}} = c(c\cup L)^+L\cup\{c\}$, with $L = \{
a^nb^n \mid n \geq 1\}$.
\end{example}

\section{Circular splicing} \label{circular}

Recall that a circular splicing system ${\cal S}= (A,I,R)$ is composed
of an alphabet $A$, an initial set $I$ of circular words, and a finite
set $R$ of rules. A rule $r=\Rul\alpha\gamma\delta\beta$ is applied to
two circular words $\cir u$ and $\cir v$, provided there exist words
$x,y$ such that $u\sim \beta x\alpha$ and $v\sim \gamma y \delta$
and produces the circular word $\cir \beta x\alpha\gamma y \delta$.

\begin{example}
  Consider the circular splicing system over $A =\{a,b\}$, with initial set
  $I = \{\cir ab\}$ and with
  the single rule $\rul abab$.  The rule expresses the fact that a
  word starting with the letter $a$ and ending with a letter $b$ can
  be inserted, in a circular word, between a letter $a$ followed by a
  letter $b$. As a consequence, the set generated by the system is the
  $\cir \{a^nb^n \mid n \geq 1\}$.
\end{example}

We now show, on this example, that  an alphabetic
circular splicing system, operating on circular words and generating a
circular language, can always be simulated  by a flat heterogeneous splicing
system. This system has the same initial set (up to full
linearization), but has an augmented set 
of rules, obtained by a kind of conjugacy of the splicing rules. To be
more precise, we introduce the following notation.
Given an alphabetic rule 
$\rul{\alpha}{\beta}{\gamma}{\delta}$, we denote by $\cir r$ the set
\begin{displaymath}
  \cir
  r=\{\rul{\alpha}{\beta}{\gamma}{\delta},\rul\delta\gamma\beta\alpha,
  \rulc\beta\alpha\gamma\delta, \rulc\gamma\delta\beta\alpha\}\,.
\end{displaymath}
The rules of the flat splicing system simulating the circular system
are the sets $\cir r$, for all rules $r$ of the circular system.
We illustrate the construction on the previous example.

\begin{example}
  Consider the flat splicing system over $A =\{a,b\}$, initial set $I
  = \{ba\}$ and with the single rule $\rul abab$.  Clearly, the rule
  cannot be applied, and consequently the language generated by the
  system reduces to $I$.

  In the world of circular
  words, the system is transformed into a heterogeneous system as follows.
  \begin{enumerate}[(1)]
  \item The initial set is now the circular class of $I$, namely the
    set $\cir I=\{ab,ba\}$.

  \item The rule $r=\rul abab$ is replaced by $\cir r$; this gives,
    by conjugacy, one new pure rule $\rul baba$ and two concatenation
    rules $\rulc abba$ and $\rulc baab$.
  \end{enumerate}

  \noindent The use of only the concatenation rules produces the set $
  \{ ab, ba, abba, baab \}$. Note that this set is not closed under
  conjugacy.
  Then, the repeated application the two pure rules
  produces the set
  \begin{displaymath}
    \{ a^{n}b^{n+m}a^{m} \mid n +m >0\} 
    \cup \{ b^{n}a^{n+m}b^{m}\mid  n +m >0\}\,.
  \end{displaymath}
  This set is now closed under conjugacy; it is the language generated
  with the four flat rules.  Moreover, it is exactly the linearization
  of the set of circular words $\cir \{a^nb^n \mid n \geq 1\}$
  generated by the circular splicing system.
\end{example}


We prove the following result which shows that the example holds in
the general case.

\begin{prop} \label{propCirc} Let ${\cal S} = (A, I, R)$ be a
  circular alphabetic splicing system, and let ${\cal S'} = (A,
  \lin{I}, R')$ be the flat heterogeneous splicing system defined by
  $R'=\bigcup_{r\in R}\cir r$. Then $\lin{\langc{{\cal
      S}}}=\langf{{\cal S'}}$.
\end{prop} 

\begin{preuve}
  We prove first the inclusion $\langc{{\cal S}}\subseteq\langf{{\cal
      S'}}$.  For this, suppose that rule
  $r=\Rul\alpha\gamma\delta\beta$ is applied, in the circular system
  ${\cal S}$ to two circular words $\cir u$ and $\cir v$. There exist
  words $x,y$ such that $u\sim \beta x\alpha$ and $v\sim \gamma y
  \delta$.  The circular word that is produced is $\cir w$ with
  $w=\beta x\alpha\gamma y \delta$.  We assume that all words in
  $\cir u,\cir v$ are in $\langf{{\cal S'}}$ and we have to show that
  any word in $\cir w$ is in $\langf{{\cal S'}}$, by the use of the
  rules in $\cir r$. First, $w$ is obtained, in ${\cal S'}$, from
  $\beta x\alpha$ and $\gamma y \delta$ by the concatenation rule
  $\rulc\beta\alpha\gamma\delta$, so $w\in \langf{{\cal S'}}$. Next,
  if $z\sim w$ and $z\ne w$, then $z=st$ and $w=ts$ for some nonempty
  words $s,t$.

  If $t$ is a prefix of $\beta x$, then there is a factorization
  $x=x'x''$ such that $t=\beta x'$, $s=x''\alpha\gamma y
  \delta$. Consequently, $z=x''\alpha\gamma y \delta\beta x'$, showing
  that $z$ is obtained, in the system ${\cal S'}$, from
  $x''\alpha\beta x'$ and $\gamma y \delta$ by the rule $r$. Since
  $x''\alpha\beta x'\sim u$, it follows that $z\in\langf{{\cal S'}}$.

  If $t=\beta x\alpha$, then $s=\gamma y \delta$ and $z=\gamma y
  \delta\beta x\alpha$. In this case, $z$ is obtained by the
  concatenation rule $\rulc\gamma\delta\beta\alpha$.

  Finally, if $\beta x\alpha\gamma$ is a prefix of $t$, then there is
  a factorization $y=y'y''$ such that $t=\beta x\alpha\gamma y'$ and
  $s=y''\delta$. Consequently, $z=y''\delta\beta x\alpha\gamma y'$,
  showing that $z$ is obtained from $y''\delta\gamma y'$ and $\beta
  x\alpha$ by the rule $\rul\delta\gamma\beta\alpha$. Since
  $y''\delta\gamma y'\sim v$, it follows again that $z\in\langf{{\cal
      S'}}$.

The converse inclusion is shown very similarly. 
\end{preuve}

The proof of the proposition relies heavily on the fact that the system is
alphabetic.

As a consequence of the proposition, we obtain the following theorem,
which is our main theorem in the circular case.

\begin{theo}\label{theo-circularContext-free}
  Let ${\cal S}=(A,I,R)$ be a circular alphabetic context-free splicing system.  Then
 $ \lin{\langc{{\cal S}}}$
is a context-free language.
\end{theo}

\begin{preuve} By Proposition~\ref{propCirc}, $\lin{\langc{{\cal
        S}}}=\langf{{\cal S'}}$, where ${\cal S'} = (A,
  \lin{I},\allowbreak R')$ is the flat heterogeneous splicing system
  defined by $R'=\bigcup_{r\in R}\cir r$.  Since $I$ is context-free,
  the language $\lin{I}$ is context-free. By
  Theorem~\ref{theoMainPlat}, the language generated by ${\cal S'}$ is
  context-free.
\end{preuve}

\bigskip

\noindent{\bf Acknowledgments:} {\em 
  We are thankful to Olivier Carton for many
  stimulating discussions during this work.  The third author also
  wishes to thank Clelia de Felice and Rosalba Zizza for inviting her
  in Salerno, for interesting discussions, and pointing out useful
  references.  }

\bibliographystyle{amsplain}
\bibliography{splicing}

\newpage
\section{Appendix: Substitution theorems for context free languages}
For sake of completeness, we give here a sketch of the proof 
Theorem \ref{subst2}, together with an example.
The proof is based on two lemmas. 
The first  deals with the case of   generalized context-free  grammar 
with a single non-terminal symbol, and the second shows how to reduce
 the number of non-terminal symbols in the general case.

\begin{lem} \label{lemsubst}
Let $G = (A, \{S\},S, R )$ be a   generalized context-free grammar 
with a single non-terminal symbol $S$. The language generated by $G$ is context-free.
\end{lem}

\begin{sketch}
Let $L$  be the language generated by $G$ and set 
$M_S = \{ m \mid S \to m \in R\}$.
Let $H  = (A \cup \{S\}, V,X, P),S \notin V$, be a usual context-free grammar 
 that generates
$M_S$. The language $L$ is generated by     the
usual context-free grammar 
$G' = (A, V\cup \{ S \},S, P \cup \{ S \to X\}  )$.
\end{sketch}

\begin{example}\label{exemple:subst}
  Let the grammar $G$ with a single non-terminal symbol $G = (A,
  \{S\}, S ,R)$ with
  \begin{displaymath}
    R= \left\{ 
      \begin{array}{lcl} 
        S&  \to & a \mid Sb^n (c^kd)^n , n \geq  1, k\geq 0 
      \end{array}
    \right \}
  \end{displaymath}
  According to the sketch of the proof given above, we define a
  grammar $H = (A \cup \{S\}, \{X, Y, Z\},X, P)$, with
  \begin{displaymath}
    P = \left\{ 
      \begin{array}{lcl} 
        X&  \to & a \mid SY\\ 
        Y&  \to & bYZ \mid bZ\\
        Z  &\to & cZ \mid d
      \end{array}
    \right.
  \end{displaymath}
  we can check that $H$ generate the language $M_S= \{ a\}\cup \{ Sb^n
  (c^*d)^n \mid n \geq 1 \}$.  we define now $G' = (A, V\cup \{ S
  \},S, P' )$ with
  \begin{displaymath}
    P' = \left\{ 
      \begin{array}{lcl}  
        S&  \to & X\\
        X&  \to & a \mid SY\\ 
        Y&  \to & bYZ \mid bZ\\
        Z  &\to & cZ \mid d
      \end{array}
    \right.
  \end{displaymath}
  The grammar $G'$ generates the same language as $G$, that is the
  language
  \begin{displaymath}
    \{a b^{n_1} (c^*d)^{n_1}  b^{n_2} (c^*d)^{n_2} \cdots b^{n_k}q
    (c^*d)^{n_k} 
    \mid k \geq 1,n_i \geq  1, 1\leq i< k\}
  \end{displaymath}
\end{example}
The  second lemma below tells us that we can reduce the problem  to
grammars with a single non-terminal symbol.

\begin{lem}
  Let $G$ be a generalized context-free grammar with at least two
  non-terminal symbols.  There is a generalized context-free grammar
  with a single non-terminal symbol which generates the same language
  as $G$ does.
\end{lem}
\begin{sketch}
  Let $G = (A, V,S, R)$ with $V$ of cardinal at least~$2$.  Let $X \in
  V$, with $X \neq S$. Define a grammar $G_X$ with one non-terminal
  symbol by $G_X = (A\cup V\setminus \{X\}, \{X\},X, R_X)$ with $R_X =
  \{ X\to m \mid X \to m \in R\}$.  Let $M_X$ the language generated
  by $G_X$. The language $M_X$ is context-free by Lemma
  \ref{lemsubst}.

  Define the substitution $\sigma_X$ on $A \cup V$ by
\begin{displaymath}
  \sigma_X(\alpha) = \left\{
    \begin{array}{l}
      M_X,   \mbox{ if } \alpha = X\\
      \{\alpha\},  \mbox{ otherwise.} \\
    \end{array}
  \right.
\end{displaymath}
Define now  the grammar $H= (A, V \setminus \{ X \}, S ,P)$ with
$P = \{v  \to \sigma_X(m) \mid v \to m \in R, v \in  V \setminus \{ X \} \}$.

The grammar  $H$ generates the same language as $G$ does, and it has a variable less than $G$.

Now it suffices to iterate the process in order to obtain a   grammar with one non-terminal symbol.
\end{sketch}

\begin{example} 
  Let $G = (A, V,S, R)$ be the generalized grammar defined by $A =
  \{a, b, c, d\}, V = \{S,T,U\}$, and
  \begin{displaymath}
    R = \left\{ \begin{array}{lcl} 
        S & \to & ST \mid a\\
        T &\to & b^n U^n, n \geq  1\\
        U  &\to & cU \mid d
      \end{array}
    \right.
  \end{displaymath}
  In the first step, we choose to remove the non-terminal $T$.
  Following the sketch of the proof given above, we define a grammar
  $G_T$ with a single non-terminal symbol $T$ by $G_T = ( A\cup\{S,
  U\}, \{T\}, T, \{T \to b^n U^n, n \geq 1\})$. Let $M_T$ be
  the language generated by $G_T$. Clearly, $M_T =\{b^n U^n \mid n
  \geq 1\}$ and $M_T$ is context-free.

  Next, we define a substitution $\sigma_T$ on $A \cup V$ by
  \begin{displaymath}
    \sigma_T(\alpha) =
    \begin{cases}
      M_T&\text{if $\alpha = T$\,,}\\
      \{\alpha\}&\text{otherwise.}
    \end{cases},
  \end{displaymath}
  and the grammar $H=(A, \{S, U\}, S ,P)$ with two non-terminal
  symbols $S,U$ by
  $P = \{v \to \sigma_T(m) \mid v
  \to m \in R, v \in V \setminus \{ T \} \}$, i.e.
  \begin{displaymath}
    P = \left\{ \begin{array}{lcl} 
        S&  \to & a \mid Sb^n U^n , n \geq  1 \\
        U  &\to & cU \mid d
      \end{array}
    \right.
  \end{displaymath}
  The grammars $H$ and $G$ generate the same language.

  To obtain a grammar with only one variable, we iterate the process
  by eliminating the variable $U$ from $H$, and we obtain the grammar
  $H' = (A, \{S\}, S ,P')$ with the unique variable $S$ and with
  \begin{displaymath}
    P '= \left\{ \begin{array}{lcl} 
        S&  \to & a \mid Sb^n (c^*d)^n , n \geq  1 
      \end{array}
    \right \}
  \end{displaymath}
  This is the grammar $G$ of  Example \ref{exemple:subst} above.
\end{example}

\end{document}